\documentclass[12pt]{iopart}
\usepackage{graphicx,epsfig,rotating}
\usepackage{amssymb}
\usepackage{url}
\newcommand{\nc}{\newcommand}
\nc{\gsim}{\mbox{\raisebox{-.6ex}{~$\stackrel{>}{\sim}$~}}}
\nc{\esim}{\mbox{\raisebox{-.6ex}{~$\stackrel{-}{\sim}$~}}}
\nc{\beq}{\begin{equation}}
\nc{\eeq}{\end{equation}}
\newcommand{\ra}{\rightarrow}

\begin{document}

\title[]{On the discovery potential of the lightest MSSM Higgs Boson at the LHC}

\author{
    R.~Kinnunen\dag, 
    S.~Lehti\dag, 
    A.~Nikitenko\ddag\footnote[3]{On leave from ITEP, Moscow, Russia} and
    P.~Salmi\dag\footnote[4]{Present address: Instituut-Lorentz, Leiden University, The Netherlands}
}

\address{\dag\ Helsinki Institute of Physics, Helsinki, Finland}
\address{\ddag\ Imperial College, University of London, London, UK}

\begin{abstract}
Production of the lightest MSSM Higgs boson h is studied at the LHC.
Isorate contours for the $\rm h \rightarrow \gamma\gamma$ and
$\rm h \rightarrow \tau^+\tau^-$ decay channels are shown in the
($\rm m_{\rm A}$,~tan$\beta$) parameter space.
Effects of the SUSY parameters, in particular the stop mixing and stop quark
mass, are investigated. Search strategies at the LHC are discussed and
the discovery potential is calculated for the CMS experiment. 
The MSSM parameter space for $\rm m_{\rm A} \gsim$~150-200~GeV/$c^2$ 
is expected to be covered with at least one decay channel with an 
integrated luminosity of 60~fb$^{-1}$. A light stop quark 
 with large stop mixing can 
affect seriously the discovery potential in the $\rm h \ra \gamma\gamma$
and $\rm h \ra \rm ZZ^*$ decay channels.

\end{abstract}





\section{Introduction}

The Minimal Supersymmetric Standard Model (MSSM) contains five Higgs bosons:
 a light CP-even Higgs boson h, a heavy 
CP-even Higgs boson H, a CP-odd Higgs boson A and two charged Higgs bosons 
H$^{\pm}$.
At tree-level the h (H) mass is bound to be below (above) the Z boson mass. 
The higher order corrections
increase this upper (lower) bound, 
the largest possible value being about 135~GeV/$c^2$ \cite{hmass}. 
The fact that in the MSSM
one Higgs boson is bound to be light gives a strong prediction 
for the mass region where the lightest Higgs boson might be seen.

The LEP and Tevatron results have already constrained the MSSM parameter
space significantly. The measurements yield lower bounds of 91.0 and
91.9~GeV/$c^2$ for the lightest CP-even Higgs boson h and for the CP-odd A,
 respectively \cite{lep_mssm}. The excluded tan$\beta$ regions are
 0.5~$<$ tan$\beta <$~2.4 for the maximal m$_{\rm h}$ scenario (maximal
mixing scenario)
and 0.7~$<$ tan$\beta <$~10.5 for the no-stop-mixing scenario~\cite{lep_mssm}, assuming top quark
mass 174.3 GeV/c$^2$. The recent results from the Tevatron, however, give the world average for 
the top mass of 178.0$\pm$4.3~GeV/$c^2$ \cite{hep-ex/0404010}. The larger top mass softens the bounds, 
for example assuming the top mass of 179.3 GeV/c$^2$ the excluded region by LEP in the maximal
mixing scenario is 0.9~$<$ tan$\beta <$~1.5 and for top mass of about 183 GeV/c$^2$ the 
exclusion vanishes \cite{new_lep}.
Some constraints have been derived from the existing data  for the other SUSY parameters. 
 The value of the trilinear coupling $\rm A_{\rm t}$ is limited to 350~GeV/$c^2$
$\lesssim \rm A_{\rm t} \lesssim$~1.5~(2.3)~TeV/$c^2$ for a light stop quark with
$\rm m_{\tilde{t}_1}$~=~200~(400)~GeV/$c^2$ and for the experimental constraint
$\rm m_{\rm h}\gsim$~90~GeV/$c^2$ \cite{hep-ph/9806315}.
The higgsino and
gaugino mass parameters $\mu$ and M$_2$ are related to
neutralino and chargino masses, and experimental mass bounds can be used to
exclude $|\mu|$ and M$_2$ values below 100~GeV/$c^2$ \cite{muM2}. 
The present experimental lower bound from LEP for the stop quark mass is 
$\sim$~100~GeV/$c^2$ \cite{lepsusy}.
 The mass limit from the Tevatron RunII with the ultimate
luminosities ($\sim$~20~fb$^{-1}$) is expected to reach m$_{\tilde{\rm t}_1} 
\sim$~240~GeV/$c^2$~\cite{fermilab}.

In this work three SUSY scenarios are considered:
a no-mixing scenario where the mixing of the left and right handed stop
eigenstates do not play any significant role and where all SUSY particles are
assumed to be heavy \cite{scenarios}, a maximal-mixing scenario which maximizes the h mass
\cite{scenarios}, 
and a light-stop scenario in which the stop quark mass is of the same order as the
top quark mass \cite{hep-ph/9806315}. These scenarios do not assume any particular
model for the soft SUSY-breaking mechanism. The stop mixing parameter is defined as X$_{\rm t}$ = 
$\rm A_{\rm t} - \mu\cot\beta$, with $\rm A_{\rm t}$ being the trilinear Higgs-stop coupling.
 The stop-mixing is maximized when X$_{\rm t}$ = $\sqrt{6} \times \rm M_{\rm SUSY}$,
 where $\rm M_{\rm SUSY}$ is the heavy SUSY scale \cite{scenarios}. In this work,
the maximal stop-mixing scenario is defined taking 
A$_{\rm t}$ = $\sqrt{6} \times \rm M_{\rm SUSY}$,
with $\rm M_{\rm SUSY}$ = 1 TeV/$c^2$.
With respect to the standard definition, this choice leads to a deviation of 
less than 1\% in the 
total production rate of $\rm pp \ra \rm h + \rm X$, $\rm h \ra \gamma\gamma$ at the LHC.
No sbottom mixing is assumed taking A$_{\rm b}$ = 0.
The higgsino mass parameter $\mu$ is set to 300~GeV/$c^2$ and the gaugino mass parameter 
M$_2$ is set to 200~GeV/$c^2$,
values chosen large enough not to be already experimentally excluded. 
All the soft SUSY breaking mass parameters are set to 1~TeV/$c^2$, and the
gluino mass M$_{\tilde{\rm g}}$ is set to 800~GeV/$c^2$.
The mass of the top quark is set to 175~GeV/$c^2$.
The values of these parameters are taken to be the same in the no-mixing scenario,
 except that of the trilinear coupling
$\rm A_{\rm t}$ which is set to zero.
For the light-stop scenario $\rm A_{\rm t}$ is taken to be 1400~GeV/$c^2$
 close to the highest possible experimentally allowed value 
($\rm A_{\rm t}$~=~1500~GeV/$c^2$) with light stop quarks \cite{hep-ph/9806315}. In this scenario
$\mu$~is set to -250~GeV/$c^2$ and  M$_2$ to 250~GeV/$c^2$. The soft SUSY breaking mass
parameters are set to 1~TeV/$c^2$ except the mass parameters of the stop sector,
which are required to be of the order of 500~GeV/$c^2$ to allow the stop quark to be light.
The actual value of the stop sector soft SUSY breaking mass
parameters vary depending on the chosen stop quark mass.

With the above values of the SUSY parameters 
 the upper bound of $\rm m_{\rm h}$ is about 127~GeV/$c^2$ in 
the maximal-mixing scenario and
about 114~GeV/$c^2$ in the no-mixing scenario.
 The sign of $\mu$ has only a small effect on the mass of the lightest 
Higgs boson.    
In the light-stop scenario with $\rm m_{\tilde{\rm t}_1}~=~200$ GeV/$c^2$, 
the upper bound of $\rm m_{\rm h}$ is 113 GeV/$c^2$, 
as in the no-mixing scenario. For $\rm m_{\tilde{\rm t}_1}~=~300$~GeV/$c^2$ 
this upper bound increases by 10~GeV/$c^2$ approaching to that of 
the maximal-mixing scenario.

The $\rm H \ra \gamma\gamma$ channel is considered one of the major discovery
channels for a light Standard Model (SM) Higgs boson and 
for the lightest scalar MSSM Higgs boson at the LHC.
There can be, however, regions of the MSSM parameter space where this discovery 
potential is reduced. 
The effect of a light stop quark in the presence of large mixing has been
calculated and the consequences for the $\rm gg \ra \rm h \ra \gamma\gamma$
channel have been discussed in Ref.~\cite{hep-ph/9806315}. An experimental study 
was performed in Ref.~\cite{NOTE2000/043} and 
the discovery potential was calculated for the $\rm h \ra \gamma\gamma$ channel in the CMS detector. 
In this earlier work, however, only the gluon-gluon fusion production process was 
simulated and other Higgs boson decay channels were not considered.
In the present work, all significant production processes are included in the calculation of the 
inclusive $\rm h \ra \gamma\gamma$ rate in the MSSM and the discovery potential
is evaluated also in the associated production and weak gauge boson 
fusion production processes $\rm qq \ra \rm qqh$. Furthermore, updated programs are 
used to calculate the cross sections and branching ratios.

 The aim of this paper is
to extend the study of the loop effects in the $\rm h \ra \gamma\gamma$  channel 
to the full discovery potential of the lightest scalar Higgs boson at the LHC.
Therefore, the $\rm h \ra \rm ZZ^* \ra \ell^+\ell^-\ell^{\prime +}\ell^{\prime -}$ and 
$\rm h/ H \ra \tau^+\tau^-$ decay channels were also studied.
The $\rm h \ra \rm ZZ^* \ra \ell^+\ell^-\ell^{\prime +}\ell^{\prime -}$ has not been
so far considered as a discovery channel in the MSSM at large tan$\beta$. It is 
shown, however, in Section 3.2 that this channel can yield a large discovery potential if the 
SUSY scenario is such that m$_{\rm h}^{\rm max} \gsim$~125~GeV/$c^2$.
 The $\rm h/ H \ra \tau^+\tau^-$ decay channels with lepton+jet
and two-lepton final states in the weak gauge boson 
fusion production have been shown to be particularly 
interesting and to cover the full of the MSSM parameter space \cite{zeppenfeld}.
In this paper the discovery potential for this channel is calculated with realistic
detector sensitivities. The CMS detector sensitivities are used and were obtained from the 
recent simulations for the discovery potential of a light SM 
 Higgs boson \cite{summary_note}.

\section{Phenomenology}
\subsection{Production cross sections}

The lightest MSSM Higgs boson h is produced through the gluon fusion 
$\rm gg \ra h$,
the associated processes $\rm q\overline{\rm q}/gg \ra \rm t \overline{\rm t}\rm h$, 
$\rm q\overline{\rm q}/gg \ra \rm b \overline{\rm b}\rm h$, $\rm qq \ra \rm Wh/Zh$
and through the weak gauge boson fusion process $\rm qq \ra \rm qqh$.
The gluon fusion process dominates 
the production over the entire parameter space.
This process is mediated by heavy quark and squark triangle loops. 
 The cross sections to the leading order (LO) and to the next to leading order 
(NLO) are calculated in this work with the program HIGLU~\cite{HIGLU}.
The top and bottom loops are included in the calculation of the Higgs boson coupling
 to gluons in this program. Since the squark loops are not included,
the decay width $\Gamma(\rm h \rightarrow \rm gg$), calculated with the HDECAY
program \cite{HDECAY}, is used to include the squark loop effects: the cross section 
given by HIGLU 
is divided by the decay width $\Gamma(\rm h \rightarrow \rm gg$)
with sparticle loops switched off, and multiplied by the decay width with all 
sparticle effects. The Higgs boson mass is kept constant in this procedure.
The corrected gluon fusion cross section with SUSY loop effects can be 
presented with the
respective branching ratios and total widths as

\begin{eqnarray}
\sigma\cdot \rm BR & = &
 \sigma( {\rm gg} \rightarrow {\rm h})
 \cdot
 \frac{\rm BR({\rm h}\rightarrow {\rm gg})^{\rm susy}}
  {\rm BR(\rm h\rightarrow \rm gg)^{\rm nosusy}}
 \frac{\Gamma_{\rm TOT}^{\rm susy}}{\Gamma_{\rm TOT}^{\rm nosusy}}
 \cdot
 \rm BR(\rm h\rightarrow\gamma\gamma)_{\rm susy},
 \label{eq:err}
\end{eqnarray}
where \verb|nosusy| refers to the branching ratio and total width calculated
assuming heavy SUSY particles and \verb|susy| to the same variables with SUSY 
spectrum determined by the given scenario. 

The cross sections for
the associated production with weak gauge bosons
$\rm qq \ra \rm Wh$ and $\rm qq \ra \rm Zh$ are calculated with the program
V2HV~\cite{spira_web} to both
leading and next to leading order. The cross sections for the production with
associated quark pairs
$\rm q\overline{\rm q}/gg \ra \rm t\overline{\rm t}\rm h$ and
$\rm q\overline{\rm q}/gg \ra \rm b\overline{\rm b}\rm h$ are calculated with the
HQQ program \cite{spira_web}, which presently includes only the LO processes.
 The cross sections for the weak gauge boson fusion $\rm qq \ra \rm qqh$
are evaluated with the VV2H program \cite{spira_web}.
The production cross sections and the contributions from the individual production
processes to the total cross section at
$\rm m_{\rm h}$~=~125.8~GeV/$c^2$ ($\rm m_{\rm A}$~=~250 GeV/$c^2$),
tan$\beta$~=~10 are shown in Table~\ref{table:crosssection}.
The gluon fusion process contributes about 80\% to the total cross 
section. This fraction is not sensitive to the Higgs boson mass and tan$\beta$.

\begin{table}[t]
\caption{Production cross sections for the lightest MSSM Higgs boson
         for $\rm m_{\rm h}$~=~125.8~GeV/$c^2$
         ($\rm m_{\rm _A}$~=~250~GeV/$c^2$) and tan$\beta$~=~10 with
         maximal stop mixing.}
\centering
\vskip 0.1 in
\small
\begin{tabular}{|c|c|c|c|c|c|c|c|}
\hline
process  & \small  gg~$\ra$~h    
         & \small qq~$\ra$~qqh 
         & \small qq~$\ra$~Wh 
         & \small qq~$\ra$~Zh 
         & \small $\rm pp \ra$~b$\overline{\rm b}$h
         & \small $\rm pp \ra$~t$\overline{\rm t}$h
         & $\sigma_{\rm TOT}$  \\
\hline
$\sigma$ (pb)  &  27.3  &  4.17  &  1.59  & 0.64 &  0.72  &  0.32 & 34.1 \\
\hline
$\sigma/\sigma_{\rm TOT}$  &  79\%  &  12\%  &  4.6\% & 1.8\% &  2.1\%  &  0.9 \% & 100 \% \\
\hline
\end{tabular}
\label{table:crosssection}
\end{table}
\vspace{2ex}

In the SM, the K-factor (defined as K = $\sigma_{\rm NLO}/\sigma_{\rm LO}$) 
for the gluon fusion process is large varying between 1.5 and 1.7 \cite{NIMB453}. 
In the MSSM, this K-factor depends on tan$\beta$ being about the same as in the SM
at small tan$\beta$ and closer to unity at large tan$\beta$
\cite{NIMB453}. The K-factors do not depend significantly on the squark mass and are 
stable against the loop effects in the gluon fusion mechanism even in the extreme situation
when one of the stop mass eigenstates is light while the other squarks are heavy and decouple
 \cite{hep-ph/9603423}.
The K-factor in the associated process qq~$\ra$~Wh is almost independent
of $\rm m_{\rm A}$ and tan$\beta$ and is about 1.3 for both the no-mixing and maximal-mixing scenario.

\subsection{Decay channels}
  
\begin{figure}[t]
  \centering
  \vskip 0.1 in
  \begin{tabular}{cc}
  \begin{minipage}{7.5cm}
    \centering
    \includegraphics[height=75mm,width=80mm]{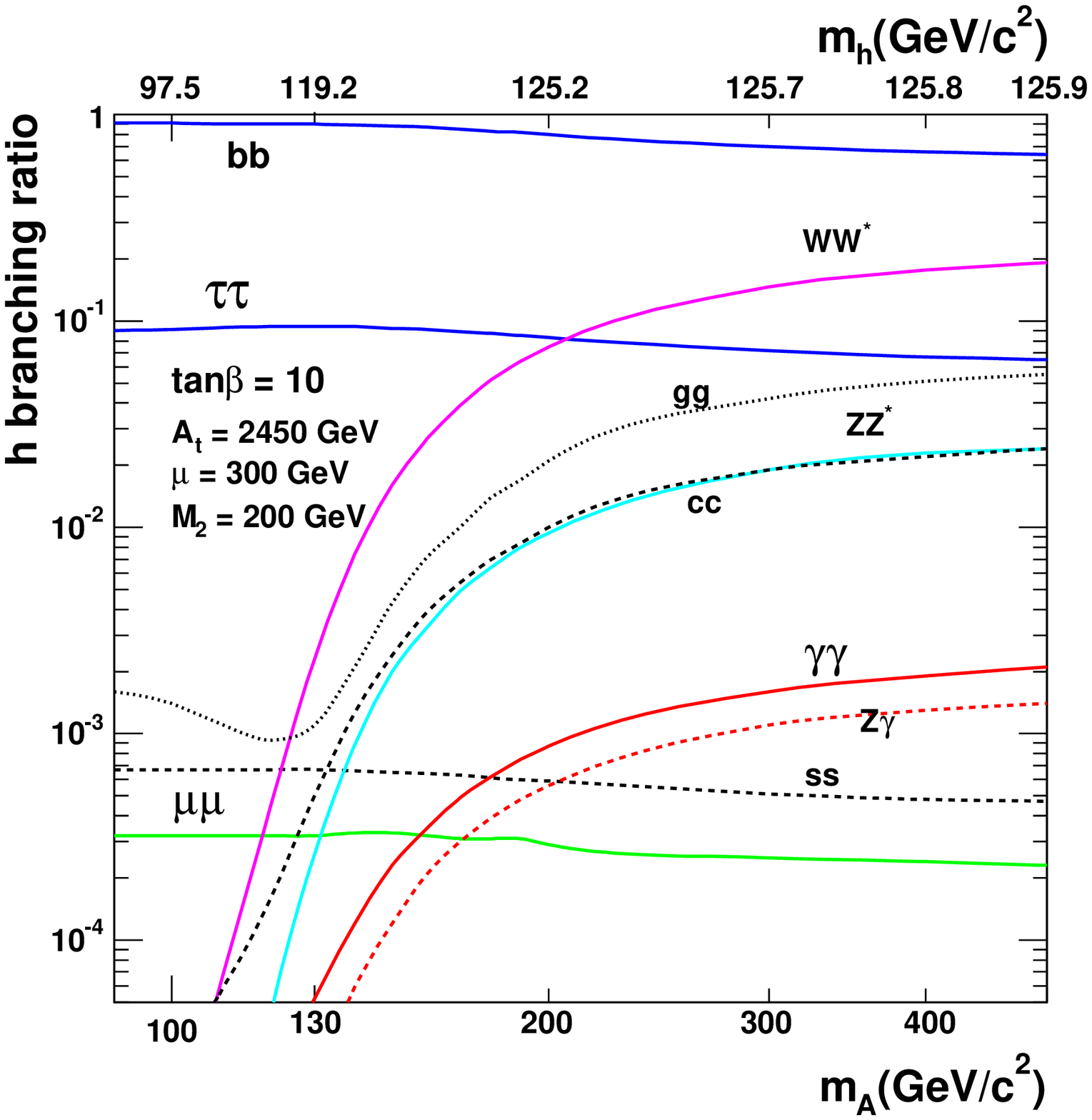} 
  \end{minipage}
  &
  \begin{minipage}{7.5cm}
    \centering
    \includegraphics[height=75mm,width=80mm]{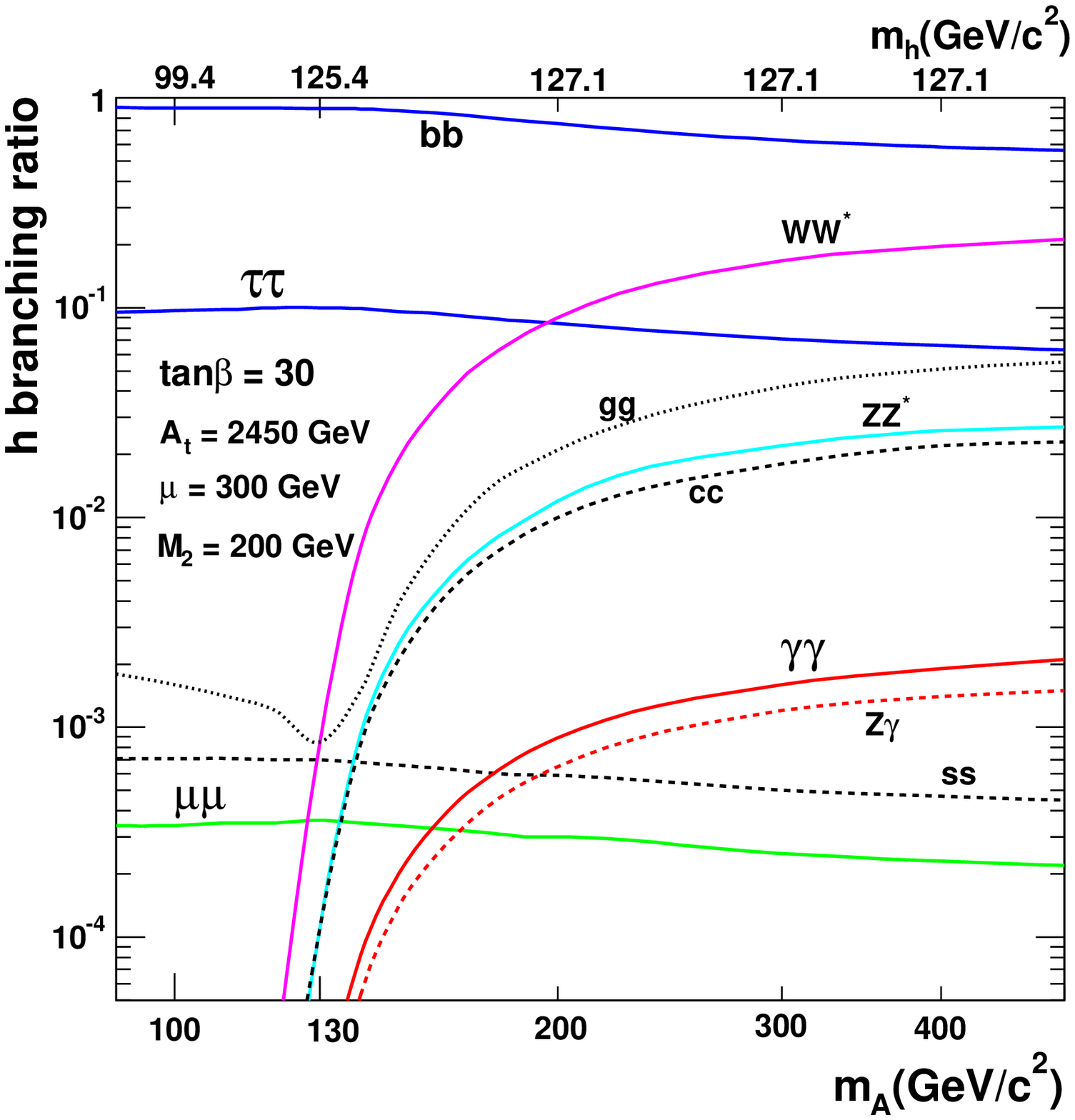}
  \end{minipage}\\
  \begin{minipage}{7.5cm}
    \centering
    \caption{Branching ratios for the lightest MSSM Higgs boson as a
             function of
             $\rm m_{\rm A}$ and  $\rm m_{\rm h}$ for tan$\beta$~=~10
             with maximal stop quark mixing.}
  \label{fig:brh_10}
  \end{minipage}
  &
  \begin{minipage}{7.5cm}
    \centering
    \caption{The same as in Fig.~\ref{fig:brh_10} but for tan$\beta$~=~30.}
    \label{fig:brh_30}
  \end{minipage}
  \end{tabular}
\end{figure}

Figures \ref{fig:brh_10} and \ref{fig:brh_30} show the branching ratios for 
the lightest MSSM 
Higgs boson as a function of $\rm m_{\rm A}$ and $\rm m_{\rm h}$ with maximal 
stop quark mixing for 
tan$\beta$ = 10 and 30, respectively. The branching ratios and decay widths are calculated with the program 
HDECAY 3.0~\cite{HDECAY}. The next to leading order (NLO) values are used for 
the decay modes throughout this study. The h~$\ra \rm b \overline{\rm b}$ decay channel dominates.
The branching ratios to weak gauge bosons h~$\ra$~ZZ$^{\ast}$ and h~$\ra$~WW$^*$ 
increase rapidly when $\rm m_{\rm h}$ approaches its maximum value reaching $\sim$ 2\% and $\sim$ 20\%, 
respectively, at large $\rm m_{\rm A}$.
For $\rm m_{\rm A}\gsim$~200~GeV/$c^2$ the
 branching ratio for the h~$\ra \gamma\gamma$ decay channel is 
between one and two per mil. The branching ratios for the h~$\ra \tau^+\tau^-$,
h~$\ra \rm b \overline{\rm b}$ and h~$\ra \mu^+\mu^-$ decay channels remain large 
also for $\rm m_{\rm A}\lesssim$~200~GeV/$c^2$, where the lightest Higgs boson
is not SM-like, due to the
enhanced couplings to the down type fermions.  

\subsection{Effect of SUSY parameters}

In the MSSM, one of the stop quarks may become much 
lighter than the other squarks if mixing between different
squark isospin eigenstates is large. The mixing can be described with the following mass 
matrix \cite{mixing_matrix}
\begin{eqnarray}
 \left(
   \begin{array}{cc}
    \rm m_{\tilde{\rm t}_{\rm L}}^2 & \rm m_{\rm top}(\rm A_{\rm t}-\mu\rm cot\beta) \\
    \rm m_{\rm top}(\rm A_{\rm t}-\mu\rm cot\beta) & \rm m_{\tilde{\rm t}_{\rm R}}^2
   \end{array}
 \right)
\end{eqnarray}
where $\tilde{\rm t}_{\rm L}$ and $\tilde{\rm t}_{\rm R}$ are the left and right handed 
eigenstates of the stop quark and 
$\rm A_{\rm t}-\mu\rm cot\beta \equiv \rm X_{\rm t}$ 
is the squark mixing 
parameter with A$_{\rm t}$ being the trilinear coupling and $\mu$ the higgsino mass 
parameter. Mixing in the third generation squark sector may be important, 
since, as seen from the off-diagonal terms of the mass matrix, 
the squark mixing is proportional to the corresponding quark mass. 
For a heavy top quark, 
mixing in the stop sector may thus produce considerable splitting between the 
mass eigenstates~$\tilde{\rm t}_{\rm 1}$,~$\tilde{\rm t}_{\rm 2}$
\begin{eqnarray}
\rm m_{\tilde{\rm t}_{\rm 1,2}}^2 & = &
\frac{1}{2}(\rm m_{\tilde{\rm t}_{\rm L}}^2+m_{\tilde{\rm t}_{\rm R}}^2)
\mp\frac{1}{2}\sqrt{(\rm m_{\tilde{\rm t}_{\rm L}}^2-
m_{\tilde{\rm t}_{\rm R}}^2)^2+
4 {\rm m}_{\rm top}^2(\rm A_{\rm t}-\mu\cot\beta)^2}
\end{eqnarray}
resulting in one very light and one very heavy stop quark. 
In above relations
${\rm m}_{\tilde{\rm t}_{\rm L}}^2 = {\rm M}_{\tilde{\rm Q}}^2+
{\rm m}_{\rm Z}^2{\rm cos}2\beta({\rm I}_3^{\tilde{\rm t}}-
{\rm e}_{\tilde{\rm t}}\sin^2\theta_{\rm W})
+{\rm m}_{\rm top}^2$ and
${\rm m}_{\tilde{\rm t}_{\rm R}}^2 = {\rm M}_{\tilde{\rm U}}^2+
{\rm m}_{\rm Z}^2\cos2\beta {\rm e}_{\tilde{\rm t}}\sin^2\theta_{\rm W}+
{\rm m}_{\rm top}^2$ 
where ${\rm M}_{\tilde{\rm Q}}$ and ${\rm M}_{\tilde{\rm U}}$ are the 
soft-SUSY breaking scalar
masses, I$_3^{\tilde{\rm t}}$ is the squark weak isospin and e$_{\tilde{\rm t}}$ the 
squark charge.
 It can be seen that in order to have 
large splitting between the two stop eigenstates and therefore a light stop 
quark, $\rm X_{\rm t}$ must be large. 
 For a common scalar mass parameter of 1~TeV, the mass of the lighter stop quark 
is of the order of 800~GeV/$c^2$. A light stop quark ($\rm m_{\rm stop}\lesssim$~300~GeV/$c^{2}$)
can be obtained  choosing the third generation scalar masses M$_{\tilde{\rm U}}$ and
 M$_{\tilde{\rm Q}}$ to be of the order of 500~GeV/$c^2$.
For these parameter values, the mass of the lightest supersymmetric 
particle LSP is of the order of 100~GeV/$c^2$, well
above the present experimental limit \cite{lep_mssm}. 
In the next sections, the lighter of the two stop mass eigenstates is denoted 
simply as a stop quark, ${\rm stop}\equiv\tilde{\rm t}_1$ and its mass $\rm m_{\rm stop}$.

\begin{figure}[h]
  \centering
  \vskip 0.1 in
  \begin{tabular}{cc}
  \begin{minipage}{7.5cm}
    \centering
    \includegraphics[height=75mm,width=80mm]{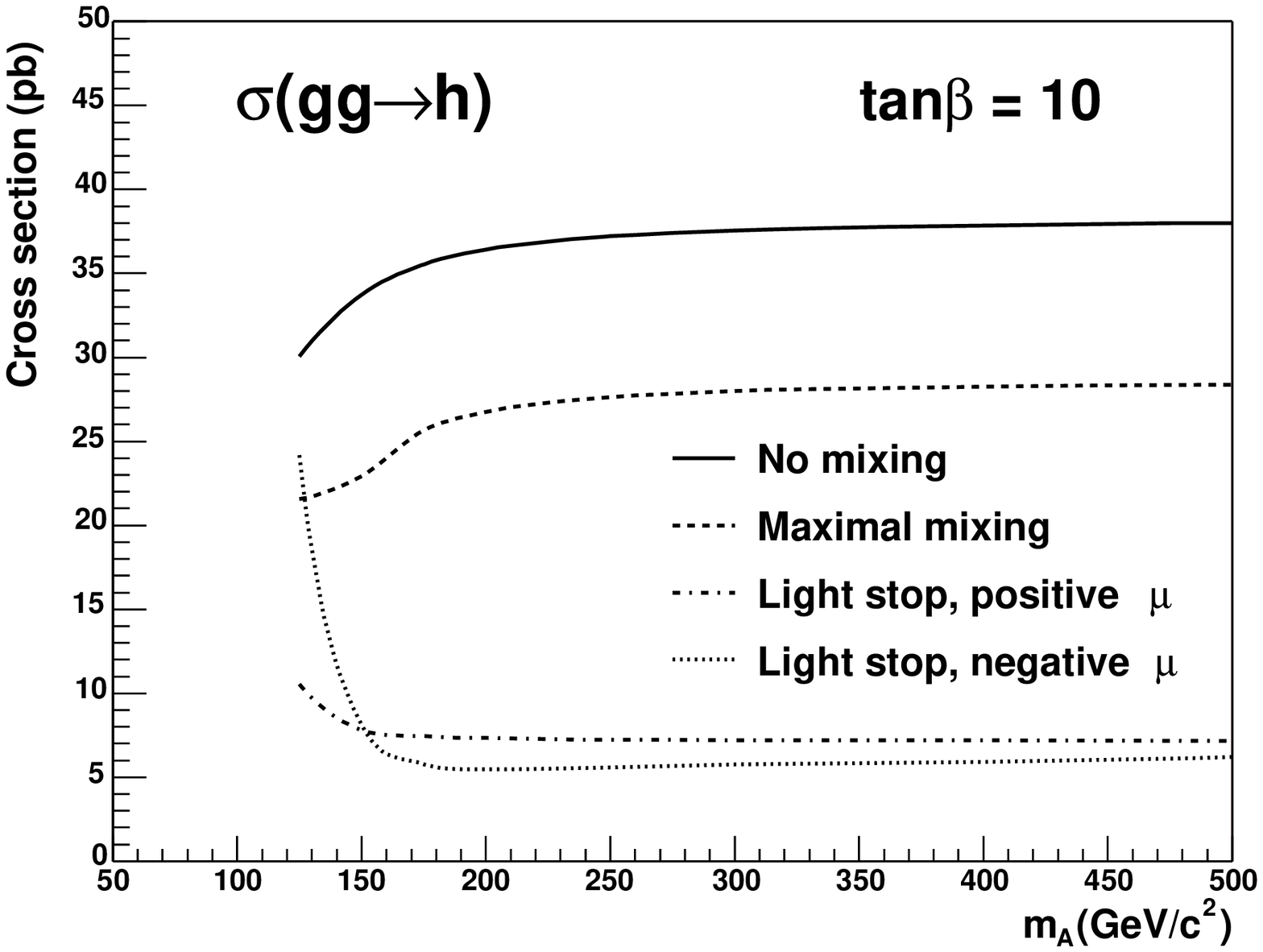}
    \caption{Cross section for the gg~$\rightarrow\rm h$ process
             with tan$\beta$~=~10 without stop mixing, with maximal stop
             mixing and with light stop quark $\rm m_{\rm stop}$~=~200 
             GeV/$c^{2}$ for $\mu<$~0 and for $\mu>$~0.}
   \label{fig:brsigma1}
  \end{minipage}
  &
  \begin{minipage}{7.5cm}
    \centering
    \includegraphics[height=75mm,width=80mm]{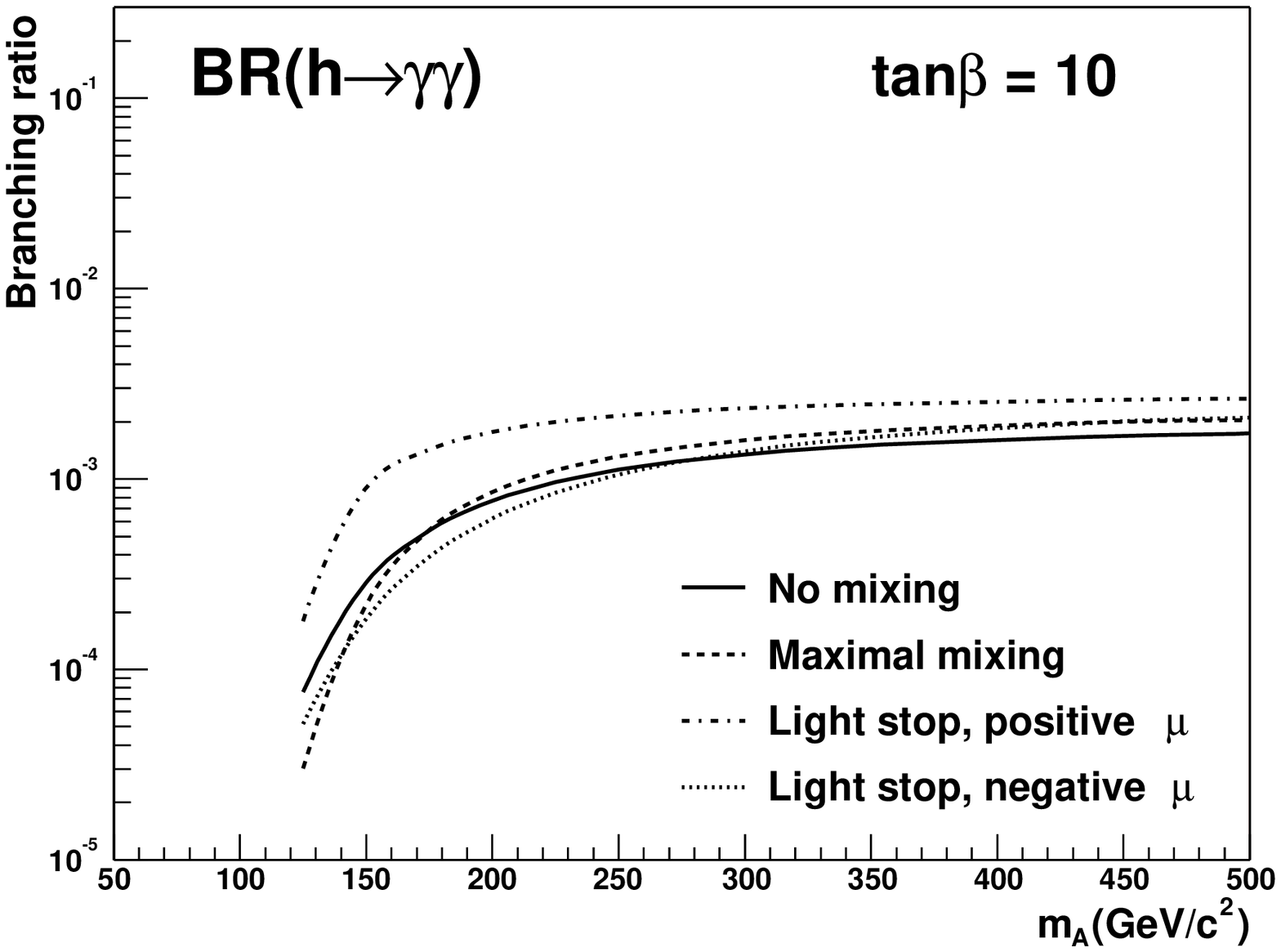}
    \caption{The branching ratio for the
             $\rm h \rightarrow\rm \gamma\gamma$ decay
             with tan$\beta$~=~10 without stop mixing, with maximal stop
             mixing and with light stop quark $\rm m_{\rm stop}$~=~200
             GeV/$c^{2}$ for $\mu<$~0 and for $\mu>$~0.}
  \label{fig:brsigma2}
  \end{minipage}
  \end{tabular}
\end{figure}

Since no supersymmetric particles have been found so far, the 
supersymmetry must be broken, the squarks do not have the same masses as 
quarks and the cancellation at loop level is less significant. 
The results from existing experiments indicate that a scenario with large mixing
in the squark (stop) sector is 
possible and more likely than a no-mixing scenario \cite{lep_mssm}.
If the mass of the stop quark is small, of the
order of the top quark mass, the cancellation effects become important.

It has been first shown in Ref.~\cite{hep-ph/9806315} that the rate 
for the gg~$\ra \rm h \ra \gamma\gamma$ process could be strongly reduced  
with large mixing
and with a light stop quark (m$_{\rm stop}\sim\rm m_{\rm top}$). 
The top - stop interference leads to a 
suppression of the top quark contribution in the loops mediating the
Higgs boson production, since the stop loop
interferes destructively with the top quark and the top and stop loops
partly cancel. The loop mediated
Higgs boson decay into photons is also affected, but since the dominant contribution comes 
now from a W loop, which interferes 
destructively with the top loop, a reduction of the top contribution by 
interfering stop loops increases the h$\ra\gamma\gamma$ partial width. 
As the W loop dominates in this partial width, the interference effect is smaller 
than in the gg~$\ra$~h process dominated by a top quark loop.
In addition to the light stop, there are  
contributions from the charged Higgs bosons, sfermions and especially from charginos,
but their net effect to the h$\ra \gamma\gamma$ partial width is small, less 
than $\sim$~10\% \cite{EurPhysJ_C1}. At large tan$\beta$ also the bottom loop contributes and even
becomes larger than the top loop contribution \cite{harlander}. 
As the reduction of the gg~$\ra$~h partial width is significantly stronger than the enhancement
of the h$\ra\gamma\gamma$ partial width, the rate
for gg~$\ra \rm h \ra \gamma\gamma$ is reduced. For m$_{\rm A}\gsim$~100~GeV/$c^2$, 
$\rm A_{\rm t}$~=~1.5~TeV/$c^2$,
m$_{\rm stop}$~=~200~GeV/$c^2$ and $\mu$~=~300~GeV/$c^2$ the rate is reduced by a factor of 
$\sim$~10 relative to the no-mixing scenario with heavy SUSY particles. The squark
sector can affect the branching ratios of the lightest Higgs boson
only via this interference phenomenon because
 the decays into gauginos are not kinematically allowed.

 Figure~\ref{fig:brsigma1} shows the cross section for 
the $\rm gg \ra \rm h$ process
and Fig.~\ref{fig:brsigma2} the branching ratios for 
the h$\ra\gamma\gamma$ decay channel corrected for
the loop effects for the following scenarios:  
no stop mixing, maximal stop mixing, light stop quark 
 $\rm m_{\rm stop}$~=~200~GeV/$c^2$ with $\mu<$~0
and light stop quark $\rm m_{\rm stop}$~=~200~GeV/$c^2$ with $\mu>$~0.  The interference 
between the stop and top quarks is clearly visible: the lighter the stop quark the 
stronger the interference and smaller the cross section. 

At large tan$\beta$ the b couplings are enhanced, but the contributions from the sbottom
loops are suppressed compared to the bottom loops 
by ($\rm m_{\rm b}/\rm m_{\tilde{b}})^2$ \cite{hep-ph/9603423}. With the LEP lower bound
for the sbottom mass \cite{lep_sbottom}, the contribution is on a per mille level.
In this work the mixing in the sbottom sector is not considered.

\subsection{Search strategies}

In the expected mass range, $\rm m_{\rm h} \lesssim$~135~GeV/$c^2$,
the lightest MSSM Higgs boson h can be searched for  
through the following decay channels: h~$\ra \gamma\gamma$, h~$\ra \gamma\rm Z$, 
 $\rm h \ra \mu^+\mu^-$, h~$\ra \rm b \overline{\rm b}$,
 $\rm h \ra \rm ZZ^* \ra \ell^+\ell^-\ell^{\prime +}\ell^{\prime -}$, 
h~$\ra$~WW$^* \ra \ell^+\ell^-\nu_{\ell}\nu_{\ell}$ and h~$\ra \tau^+\tau^-$. The searches in the
 $\rm h \ra \gamma\gamma$, $\rm h \ra \mu^+\mu^-$ and 
$\rm h \ra \rm ZZ^* \ra \ell^+\ell^-\ell^{\prime +}\ell^{\prime -}$ 
channels are based on the small total width of the
Higgs boson in this mass range (in the SM and MSSM) exploiting the precise photon energy and lepton momentum
measurements for the Higgs boson mass reconstruction 
\cite{tdr:ecal,tdr:tracker,tdr:muon}. 
The $\rm h \ra \gamma\gamma$ and 
$\rm h \ra \rm ZZ^* \ra \ell^+\ell^-\ell^{\prime +}\ell^{\prime -}$ 
channels are expected to yield their largest reaches in the inclusive production,
dominated by the gluon fusion process.
The $\rm h \ra \gamma\gamma$ channel can be searched for also in
the associated production processes $\rm t\overline{\rm t}\rm h$ and 
$\rm Wh$ with a requirement of an isolated lepton 
from the $\rm W \ra \ell\nu_{\ell}$ decay \cite{lgamma}. The 
signal-to-background ratios are larger but the event rates are
smaller than for the inclusive production. 
For the $\rm h \ra \rm b \overline{\rm b}$ decay channel, suppression of the 
QCD multi-jet background is possible only in the associated production 
processes $\rm t \overline{\rm t}\rm h$ and Wh with a requirement of an
isolated lepton from the $\rm W \ra \ell\nu_{\ell}$ decay \cite{volker}.  
The $\rm h \ra \gamma\gamma$, $\rm h \ra \mu^+\mu^-$,
$\rm h \ra \rm WW^* \ra \ell^+\ell^-\nu_{\ell}\nu_{\ell}$, $\rm h \ra \tau^+\tau^-$ and possibly 
h~$\ra \rm b \overline{\rm b}$ decay channels
can be searched for also in the weak gauge boson fusion production process 
$\rm qq \ra \rm qqh$. In this production mechanism tagging of the forward jets and vetoing
on central hadronic jets can be used to efficiently suppress the QCD multi-jet, W+jets and 
$\rm t\overline{\rm t}$ backgrounds \cite{sasha}.
The $\rm h \ra \tau^+\tau^-$ channel is particularly interesting in the MSSM as the couplings to
down type fermions are tan$\beta$ enhanced relative to SM couplings. Due to the tiny branching
ratios, the h~$\ra \gamma\rm Z$ and 
 $\rm h \ra \mu^+\mu^-$ decay channels may be exploited only with the integrated luminosities 
exceeding 100~$\rm fb^{-1}$.

\vspace{ 3mm}
\section{Inclusive production channels}
\subsection{$\boldmath{\rm h \ra \gamma\gamma}$}

\begin{figure}[h]
  \centering
  \vskip 0.1 in
  \begin{tabular}{cc}
  \begin{minipage}{7.5cm}
    \centering
    \includegraphics[height=75mm,width=80mm]{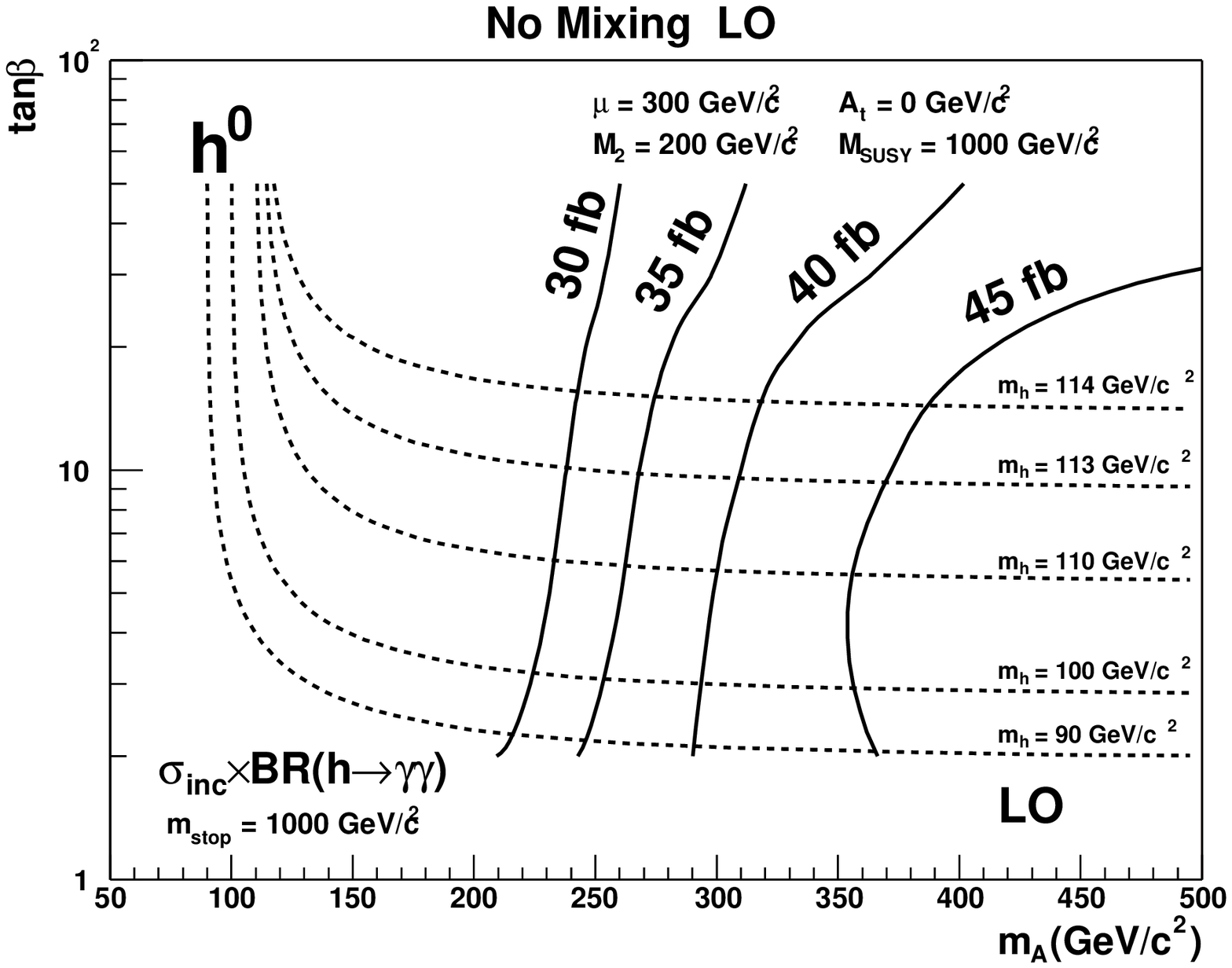}
    \caption{Isorate (cross section times branching ratio) curves for
             the inclusive $\rm h \ra \gamma\gamma$ channel in the
             no-mixing scenario with LO cross sections. The
             isomass curves for the lightest MSSM Higgs boson are shown with
             dashed lines.}
    \label{fig:nomix_gamma_lo}
  \end{minipage}
  &
  \begin{minipage}{7.5cm}
    \centering
    \includegraphics[height=75mm,width=80mm]{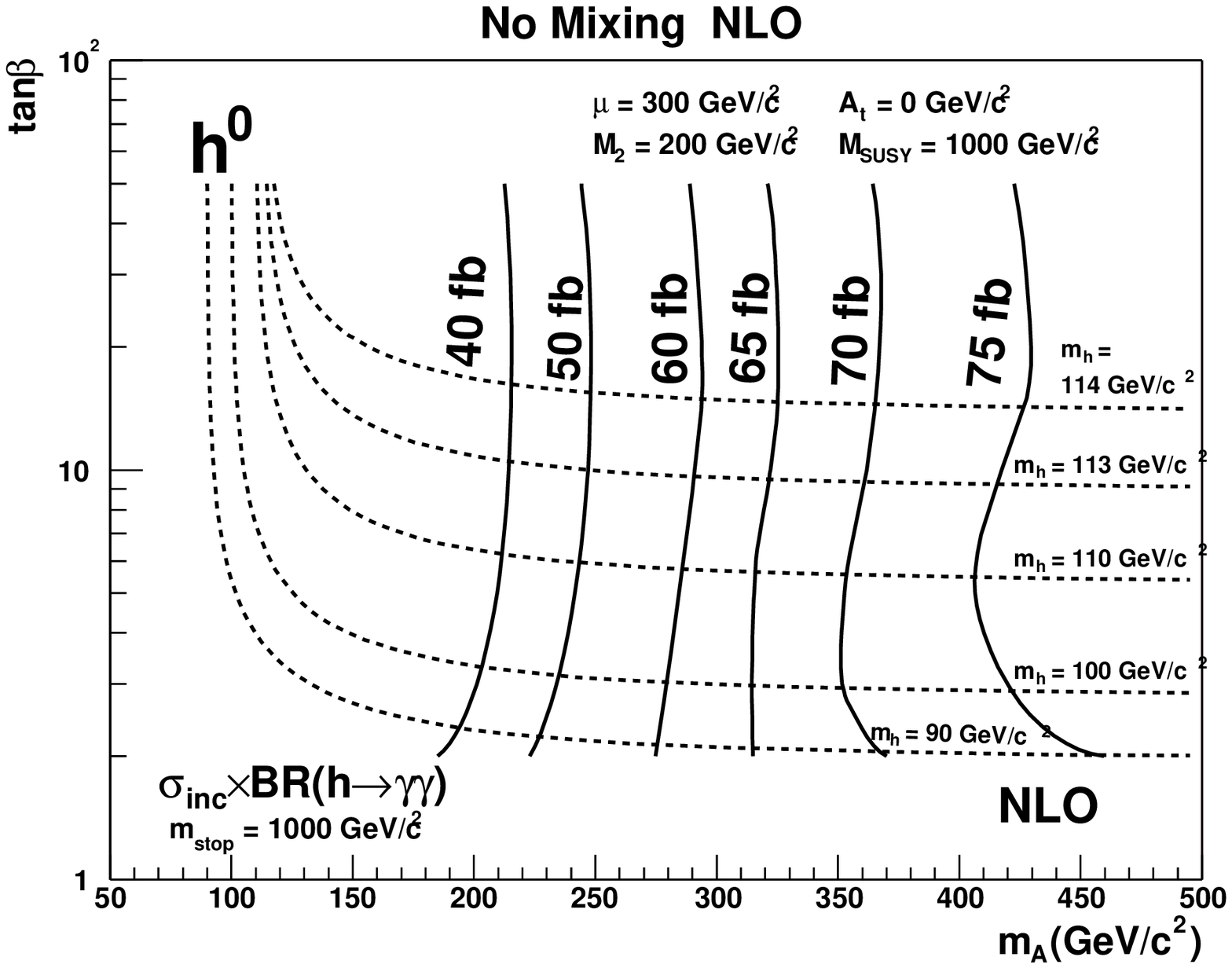}
    \caption{Isorate (cross section times branching ratio) curves for
             the inclusive $\rm h \ra \gamma\gamma$ channel in the 
             no-mixing scenario with NLO cross sections. The
             isomass curves for the lightest MSSM Higgs boson are shown with
             dashed lines.}
    \label{fig:nomix_gamma_nlo}
  \end{minipage}
  \end{tabular}
  \vskip 0.1 in

  \centering
  \begin{tabular}{cc}
  \begin{minipage}{7.5cm}
    \centering
    \includegraphics[height=75mm,width=80mm]{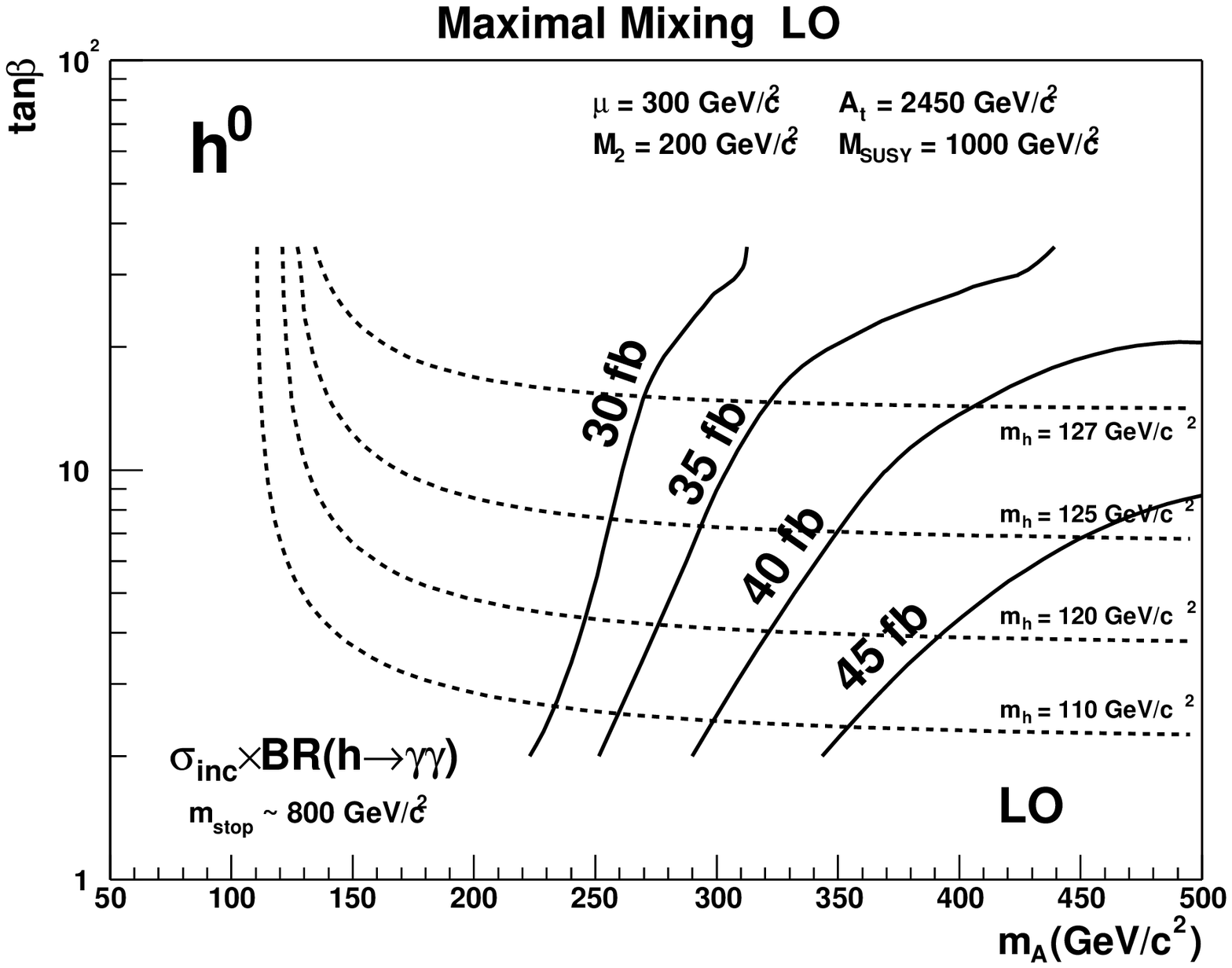}
  \caption{Isorate curves for
           the inclusive $\rm h \ra \gamma\gamma$ channel in the 
           maximal-mixing scenario with LO cross sections. }
  \label{fig:maxmix_gamma_lo}
  \end{minipage}
  &
  \begin{minipage}{7.5cm}
    \centering
    \includegraphics[height=75mm,width=80mm]{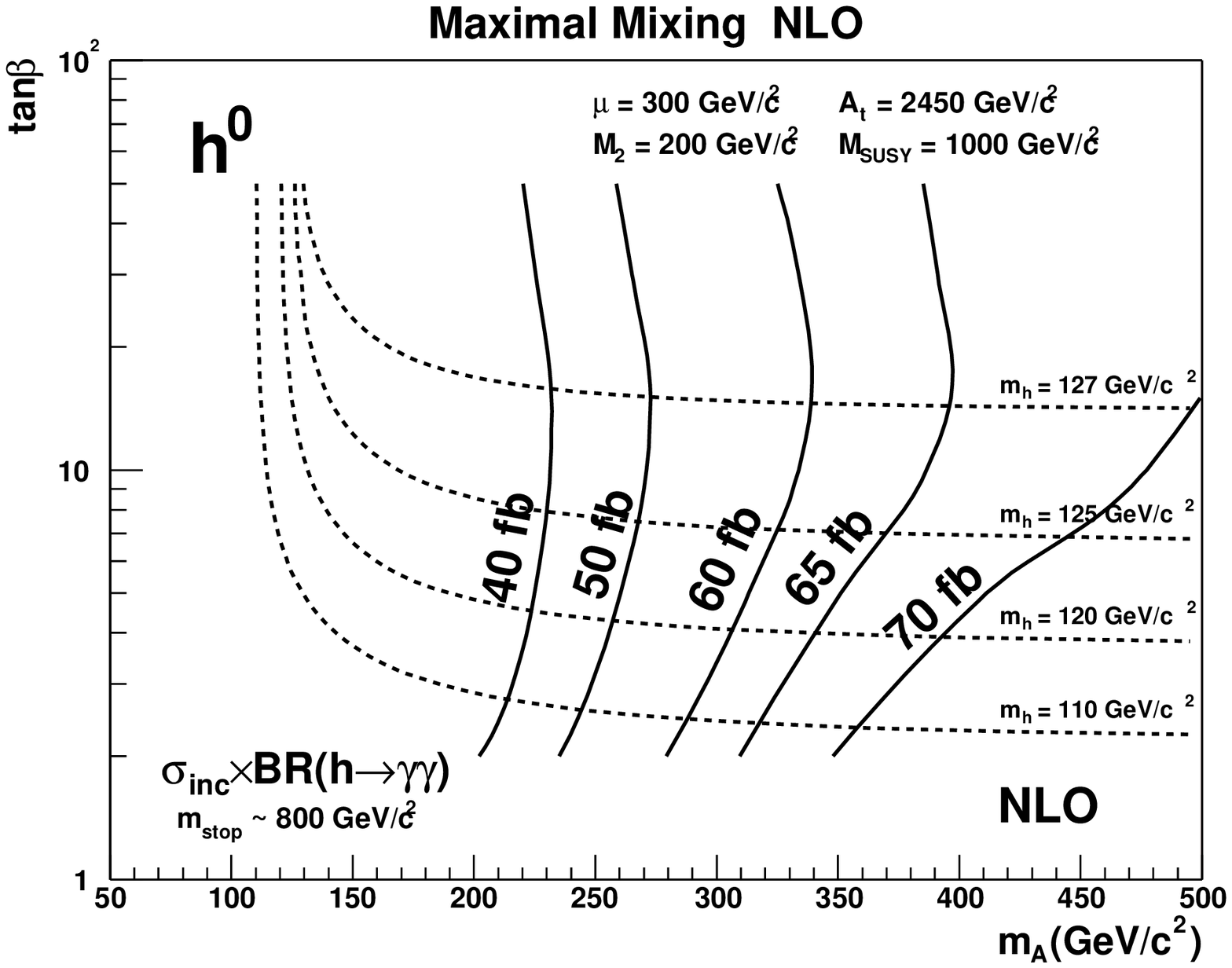}
  \caption{Isorate curves for
           the inclusive $\rm h \ra \gamma\gamma$ channel in the
           maximal-mixing scenario with NLO cross sections. }
  \label{fig:maxmix_gamma_nlo}
  \end{minipage}
  \end{tabular}

\end{figure}

The isorate (cross section times branching ratio) 
curves for the $\rm h \ra \gamma\gamma$ channel in the inclusive production 
in the no-mixing scenario are shown in
Fig.~\ref{fig:nomix_gamma_lo} with LO cross sections and in Fig.~\ref{fig:nomix_gamma_nlo} with
NLO cross sections. The isorate curves for the inclusive production in the maximal-mixing scenario are 
shown in Fig.~\ref{fig:maxmix_gamma_lo} with LO and in Fig.~\ref{fig:maxmix_gamma_nlo} 
with NLO cross sections. Due to the 
 larger Higgs boson mass $\rm m_{\rm h}$ in the maximal-mixing
 scenario for fixed $\rm m_{\rm A}$ and tan$\beta$ 
the cross section is smaller than that in the no-mixing scenario. This
decrease is compensated by a larger $\rm h \ra \gamma \gamma$  branching ratio
resulting in a $\sim$~3\% lower production (cross section times branching ratio) 
rate relative to the no-mixing scenario.
 Although in this scenario the stop quark is rather heavy 
$\rm m_{\rm stop} \simeq 800$~GeV/$c^2$, the effect of the
 virtual stop loops suppresses the cross section by
approximately 10\% relative to the no-mixing scenario. A negative higgsino 
mass parameter would yield a further small suppression.

\begin{figure}[t]

 \centering
  \vskip 0.1 in
  \begin{tabular}{cc}
  \begin{minipage}{7.5cm}
  \includegraphics[height=75mm,width=80mm]{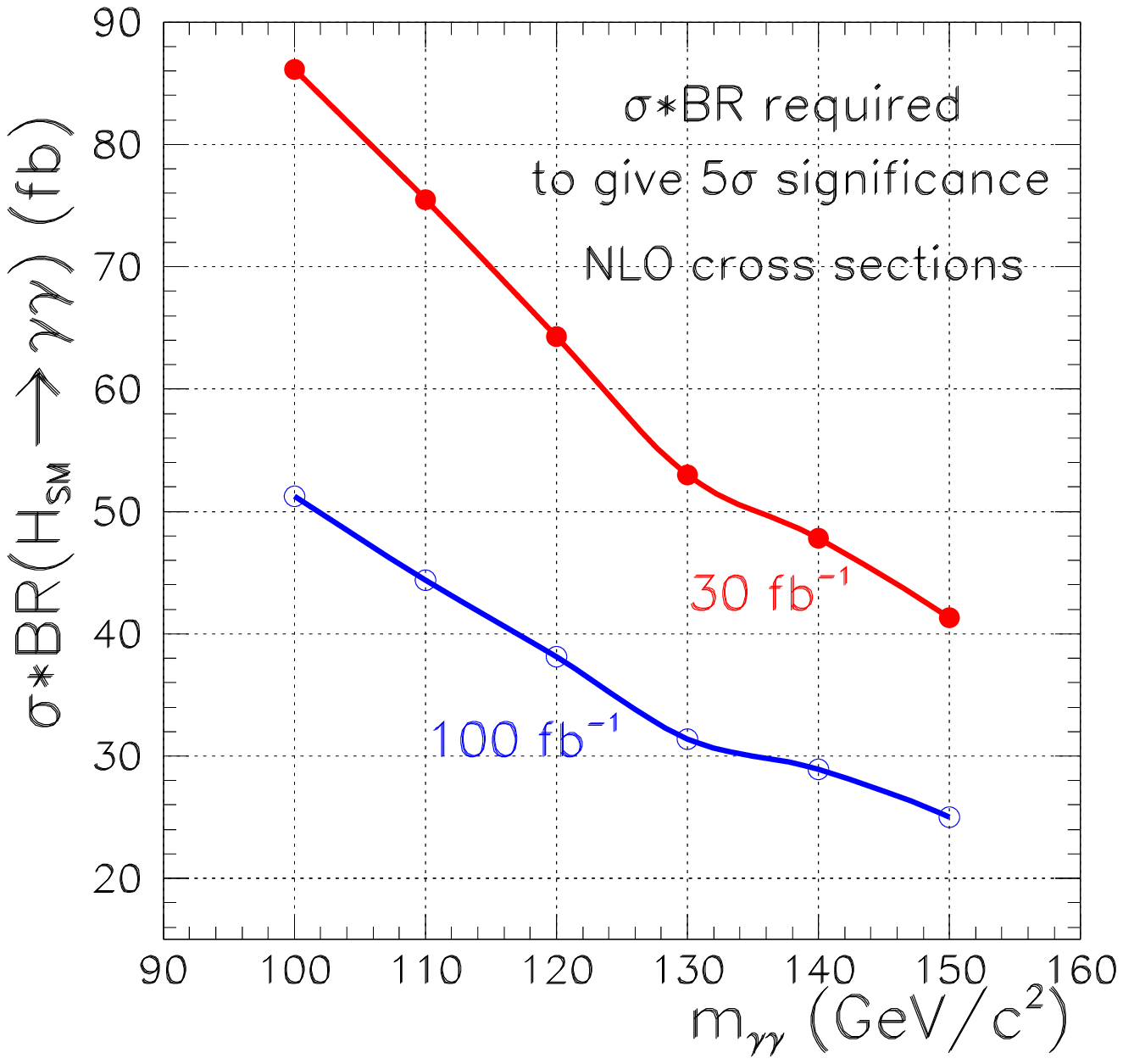}
  \end{minipage}
 &
\begin{minipage}{7.5cm}
  \centering
  \includegraphics[height=75mm,width=80mm]{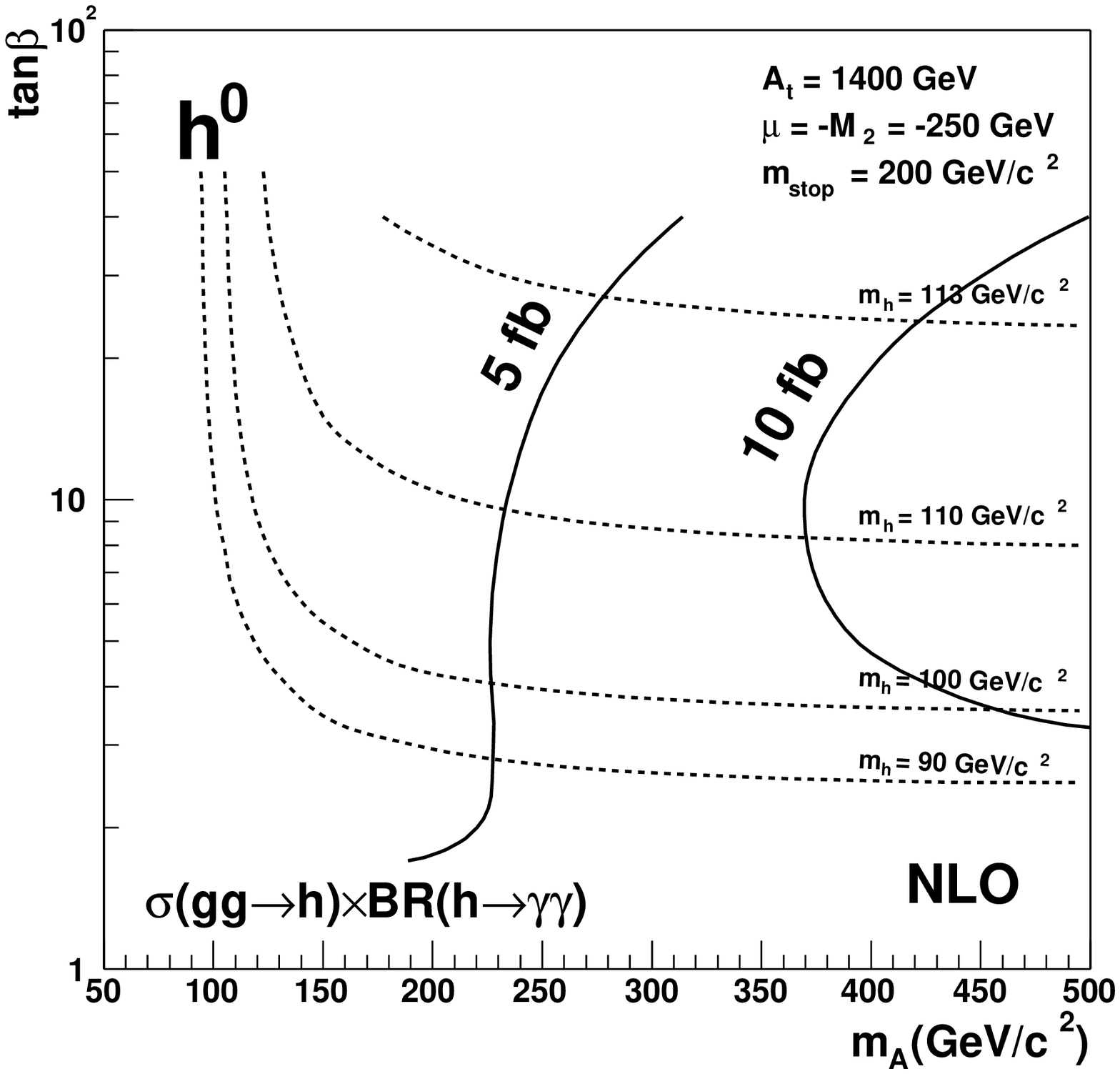}
  \end{minipage}
\\
  \begin{minipage}{7.5cm}
  \caption{Cross section times branching ratio to give a
           5$\sigma$ statistical significance
           for the inclusive $\rm H \ra \gamma\gamma$ channel in the SM
           for 30 and 100 fb$^{-1}$ assuming NLO
           cross sections for the backgrounds \cite{tdr:ecal}.}
  \label{fig:expcurves}
  \end{minipage}
 &
\begin{minipage}{7.5cm}
  \centering
  \caption{Isorate curves for the $\rm gg \ra \rm h \ra \gamma\gamma$ channel
      with a light stop quark $\rm m_{\rm stop}~=~200$~GeV/$c^2$
      with LO cross sections.}
  \label{fig:gfusion_stop200}
  \end{minipage}
  \end{tabular}

  \begin{tabular}{cc}
  \begin{minipage}{7.5cm}
    \centering
    \includegraphics[height=75mm,width=80mm]{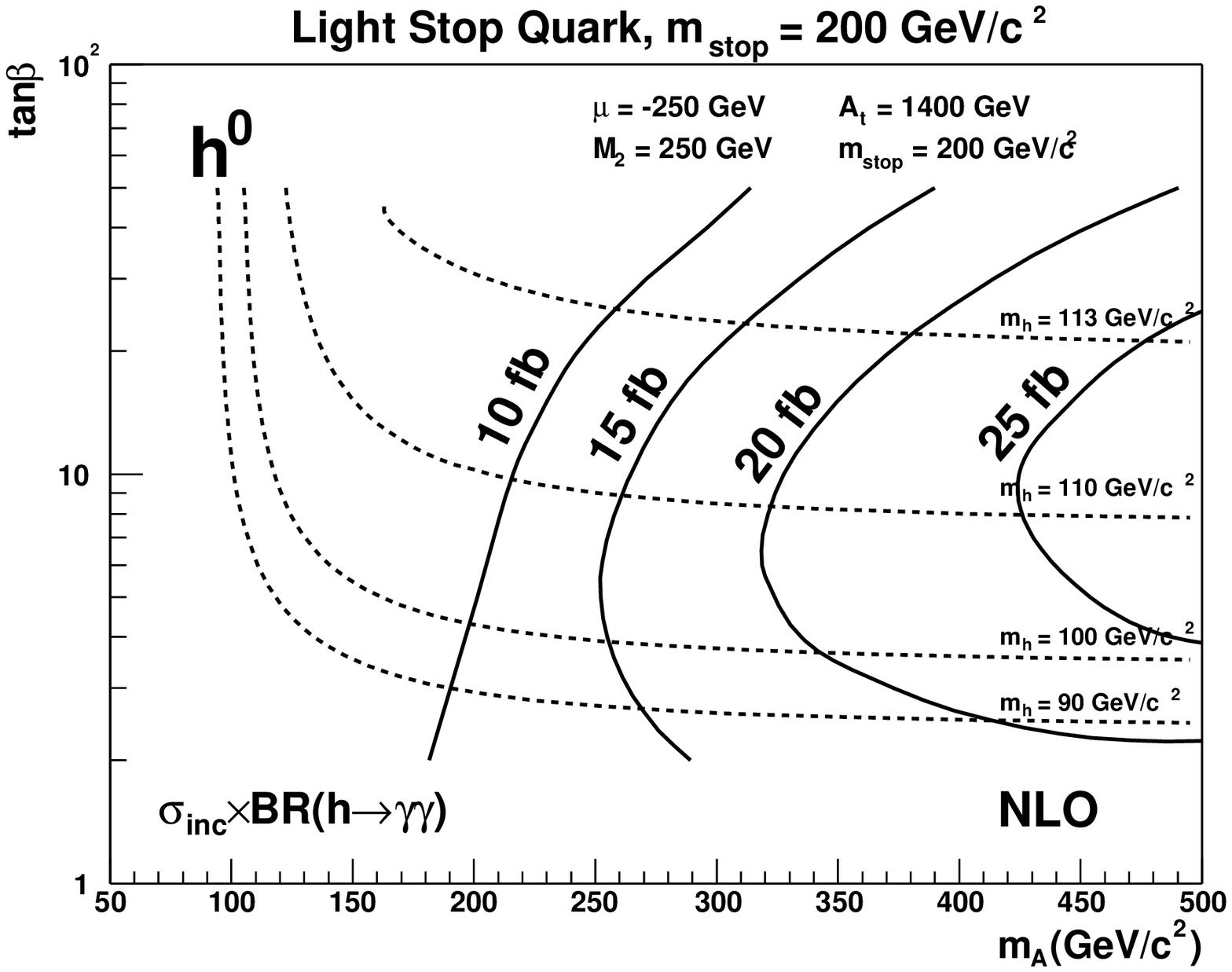}
  \caption{Isorate curves for the inclusive $\rm h \ra \gamma\gamma$ channel
           with a light stop quark $\rm m_{\rm stop}$~=~200~GeV/$c^2$ with NLO cross sections.}
  \label{fig:inc_stop200}
  \end{minipage}
  &
  \begin{minipage}{7.5cm}
    \centering
    \includegraphics[height=75mm,width=80mm]{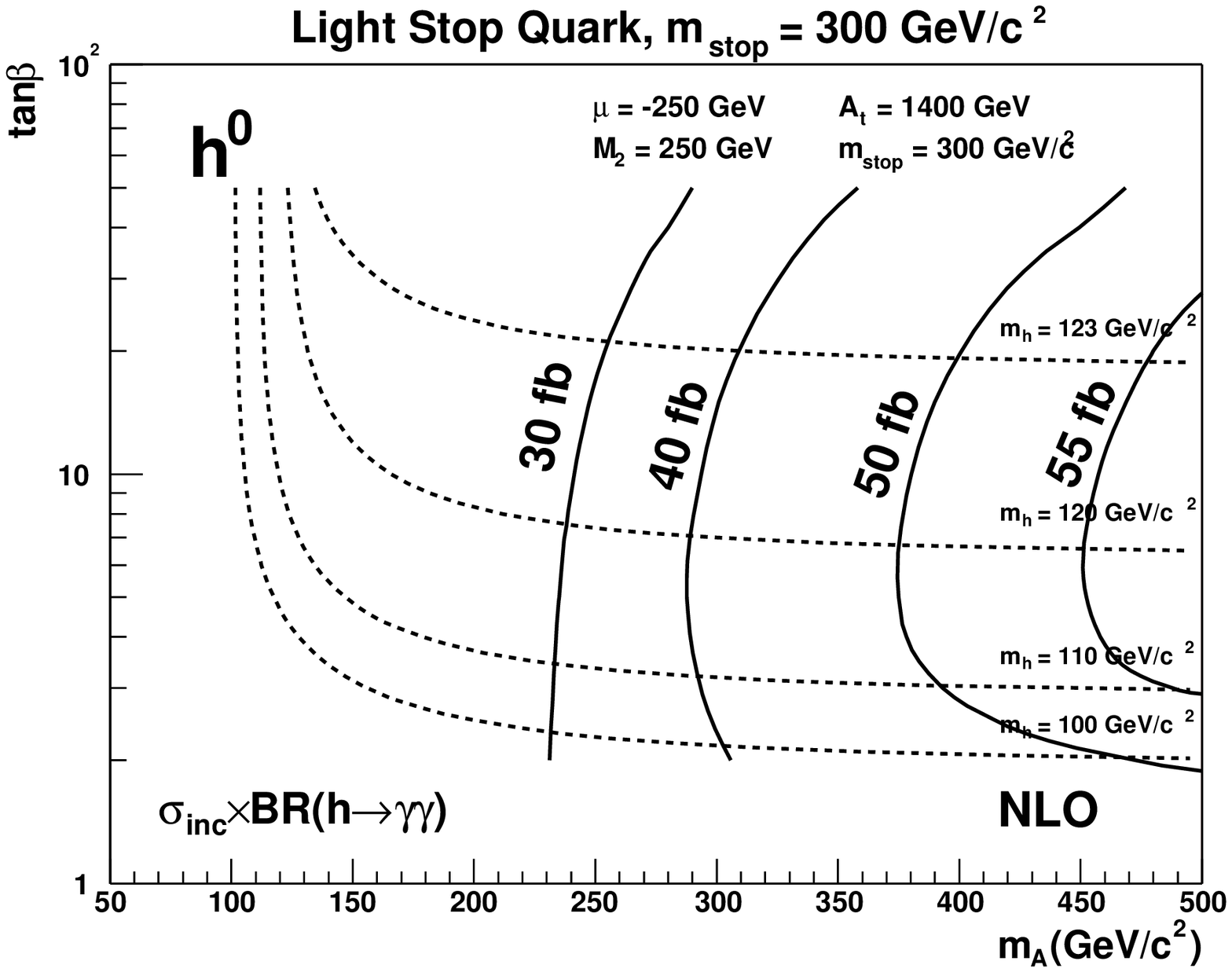}
  \caption{Isorate curves for the inclusive $\rm h \ra \gamma\gamma$ channel
           with a light stop quark $\rm m_{\rm stop}$~=~300~GeV/$c^2$ with NLO cross sections.}
  \label{fig:inc_stop300}
  \end{minipage}
  \end{tabular}
\end{figure}

Figure \ref{fig:expcurves} shows the cross section times branching ratio
required for a 5$\sigma$
statistical significance in the inclusive $\rm h \ra \gamma\gamma$ channel as a 
function of the invariant two-photon mass for 30 and 100~$\rm fb^{-1}$
 in the CMS detector \cite{tdr:ecal}.
The NLO cross
sections are assumed for the signal and backgrounds. 
In the mass range of the lightest MSSM Higgs boson, 
$\rm m_{\rm h} \lesssim$~127~GeV/$c^2$, production rates at least 55 and 33~fb are needed 
 to obtain a 5$\sigma$ statistical significance with an integrated luminosities of 30 and 100~$\rm fb^{-1}$,
respectively. 
In the no-mixing scenario with $\rm m_{\rm h} \lesssim$~114~GeV/$c^2$ the minimal production rates 
required for these luminosities are 71 and 42~fb, respectively.
Larger rates are needed at lower mass values due to the increasing backgrounds.
With these detector sensitivities a 5$\sigma$-discovery potential is expected 
for $\rm m_{\rm A} \gsim$~200 and 300~GeV/$c^2$ 
with integrated luminosities of 100 and 30~$\rm fb^{-1}$, respectively. The reach 
is approximately the same in the no-mixing and maximal-mixing scenario.

 Figures \ref{fig:gfusion_stop200}, \ref{fig:inc_stop200} and 
\ref{fig:inc_stop300} show the isorate curves 
for the $\rm h \ra \gamma\gamma$ channel in the light-stop
scenario. 
Figure \ref{fig:gfusion_stop200} shows the isorate curves for the
dominating gluon fusion process which is affected most by a light
stop quark. 
Figure~\ref{fig:inc_stop200} shows the isorate curves  
in the inclusive 
production for a very light stop quark, m$_{\rm stop}~=~200$ GeV/$c^2$. 
Since the gluon fusion process is the dominating production mechanism, 
the effect of a light stop on the inclusive production is large, too.
A discovery in the inclusive $\rm h \ra \gamma\gamma$ channel with such a 
light stop quark
could be possible only for $\rm m_{\rm A} \gg $~500 GeV/$c^2$
for integrated luminosities exceeding 100~$\rm fb^{-1}$.
Figure \ref{fig:inc_stop300} shows the isorate curves for 
the inclusive production with m$_{\rm stop}~=~300$ GeV/$c^2$. 
For this value of m$_{\rm stop}$ a discovery is possible with 100~$\rm fb^{-1}$ 
in part of the parameter space for about 
$\rm m_{\rm A} \gsim$~400~GeV/$c^2$ and 
tan$\beta \gsim$~10. For $\rm m_{\rm stop}\gtrsim$~400~GeV/$c^2$ the 
interference
effect is already small and the rate is close to that of the no-mixing and
maximal-mixing scenarios.


\subsection{$\boldmath{\rm h \ra \rm ZZ^* \ra \ell^+\ell^-\ell^{\prime +}\ell^{\prime -}}$}

The four-lepton channel, the $\rm H \ra \rm ZZ^* \ra \ell^+\ell^-\ell^{\prime +}\ell^{\prime -}$,
has been shown to be the major discovery channel over the large mass rangion in the SM \cite{summary_note}.
In the MSSM, the heavier scalar H could be searched for in the four-lepton channel at small tan$\beta$.
For $\rm m_{\rm H}\lesssim 2 \rm m_{\rm Z}$, where the detector resolution dominates, the 
discovery potential could be obtained from that for the SM Higgs boson while for 
$\rm m_{\rm H}\gsim 2 \rm m_{\rm Z}$ dedicated studies are needed due to the difference
of total Higgs boson widths between the SM and MSSM in this region. For the lighter
scalar h, a discovery could be possible close to the maximal possible value of $\rm m_{\rm h}$ at 
large tan$\beta$ and  $\rm m_{\rm A}$.
The discovery potential is strongly dependent on the lowest possible mass 
value accessible in the (pure) SM scenario, due to the fast decreasing $\rm h \ra \rm ZZ^*$ branching ratio.
The CMS studies have shown that this value could be as low as $\rm m_{\rm H} \sim$~120~GeV/$c^2$
with an integrated luminosity of 100~$\rm fb^{-1}$ combining the electron and muon channels 
\cite{4lepton,lassila}. Therefore a significant region at large tan$\beta$ could be covered
in the maximal-mixing scenario while  no sensitivity is possible in the MSSM
in the scenarios where the mass of the lighter scalar is below $\sim$~120~GeV/$c^2$.

\section{Associated production channels}

\begin{figure}[h]
  \centering
  \vskip 0.1 in
  \begin{tabular}{cc}
  \begin{minipage}{7.5cm}
    \centering
    \includegraphics[height=75mm,width=80mm]{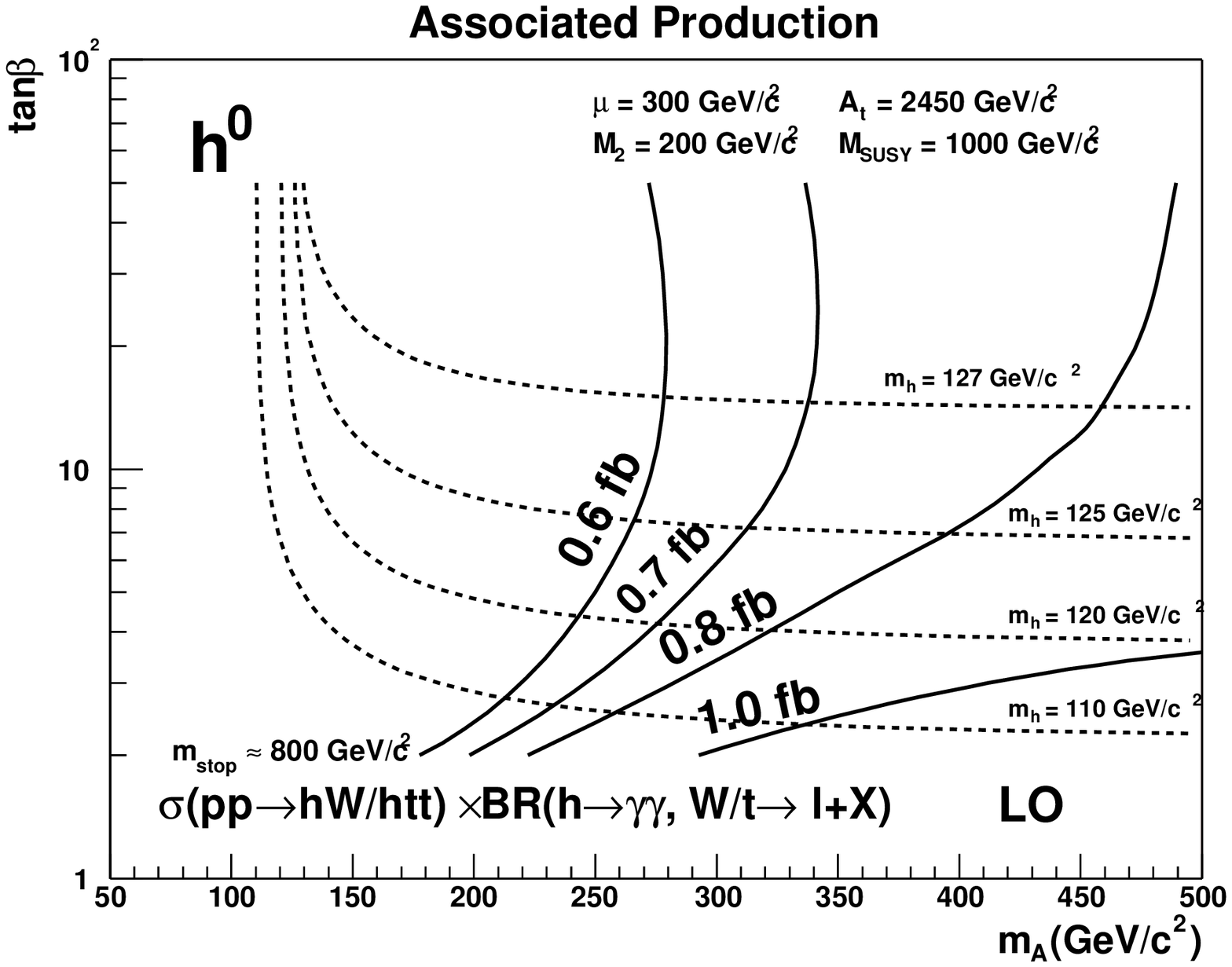}
  \end{minipage}
  &
  \begin{minipage}{7.5cm}
    \centering
    \includegraphics[height=75mm,width=80mm]{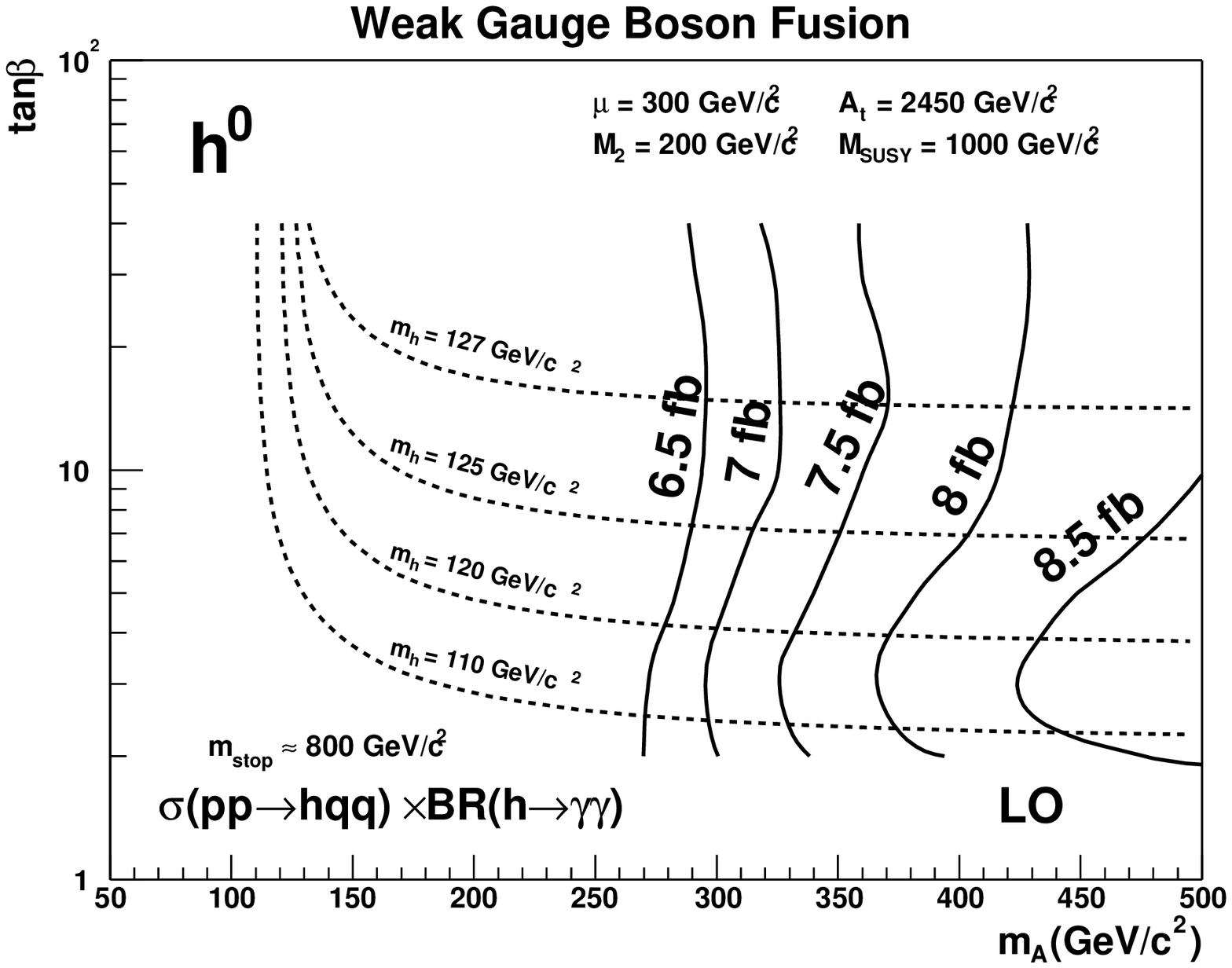}
  \end{minipage}\\
  \begin{minipage}{7.5cm}
    \centering
    \caption{Isorate curves for h$\rightarrow\gamma\gamma$ in the
             associated production processes qq~$\ra$~Wh and
             $\rm q\overline{\rm q}/\rm gg \ra \rm t \overline{\rm t}h$
             with $\gamma\gamma\ell$ final states. Maximal stop
             mixing and LO cross sections assumed.}
  \label{fig:associated_maxmix_lo}
  \end{minipage}
  &
  \begin{minipage}{7.5cm}
    \centering
    \caption{Isorate curves for weak gauge boson fusion qq~$\ra$~qqh,
             h~$ \ra \gamma\gamma$ with maximal stop mixing.}
  \label{fig:wwhgg}
  \end{minipage}
  \end{tabular}
\end{figure}

The isorate curves for the $\rm h \ra \gamma\gamma$ channel in the associated production
combining the $\rm qq \ra \rm Wh$ and $\rm q\overline{\rm q}/ \rm gg \ra \rm t \overline{\rm t}\rm h$ processes 
are shown in Fig.~\ref{fig:associated_maxmix_lo}
in the maximal-mixing scenario. The branching ratio for the $\rm W \ra \ell\nu_{\ell}$ decay is included. 
The $\rm qq \ra \rm Wh$ process dominates the production and is large at small tan$\beta$, enhancing the 
total rate in this region. 
 The cross section of the associated 
production is not sensitive to the mixing and stop mass effects. The production rate  
can be only affected through the loop mediated $\rm h \ra \gamma\gamma$ decay process.
 
 In the SM, the $\rm H \ra \gamma\gamma$ decay channel has been shown to be
accessible in the associated 
 $\rm qq \ra \rm WH$ production process with an integrated luminosity of 
100~$\rm fb^{-1}$ \cite{lgamma}.
The total production rate required for a 5$\sigma$ statistical significance 
 in the $\rm H \ra \gamma\gamma$ decay channel is between 0.8 and 0.6~fb 
for 110~$<\rm m_{\rm H}<$~127~GeV/$c^2$. In the MSSM such a rate is expected
only in the region of large $\rm m_{\rm A}$ and small tan$\beta$ as can be seen from 
Fig.~\ref{fig:associated_maxmix_lo}. The
 $\rm H \ra \rm b\overline{\rm b}$ decay channel has been investigated in the associated 
$\rm q\overline{\rm q} \ra \rm t \overline{\rm t}\rm H$ process \cite{volker}.
A 5$\sigma$ statistical significance is reached in the SM with 
integrated luminosities exceeding 40~$\rm fb^{-1}$ around $\rm m_{\rm H} \sim $~120~GeV/$c^2$ \cite{volker}.
Due to the enhanced  $\rm h \ra \rm b\overline{\rm b}$ couplings at large tan$\beta$,
a significant region has been shown to be covered with this decay channel in the MSSM \cite{volker2}.

\section{Weak gauge boson fusion production channels}

\begin{figure}[h]
  \centering
  \vskip 0.1 in
  \begin{tabular}{cc}
  \begin{minipage}{7.5cm}
    \centering
    \includegraphics[height=75mm,width=80mm]{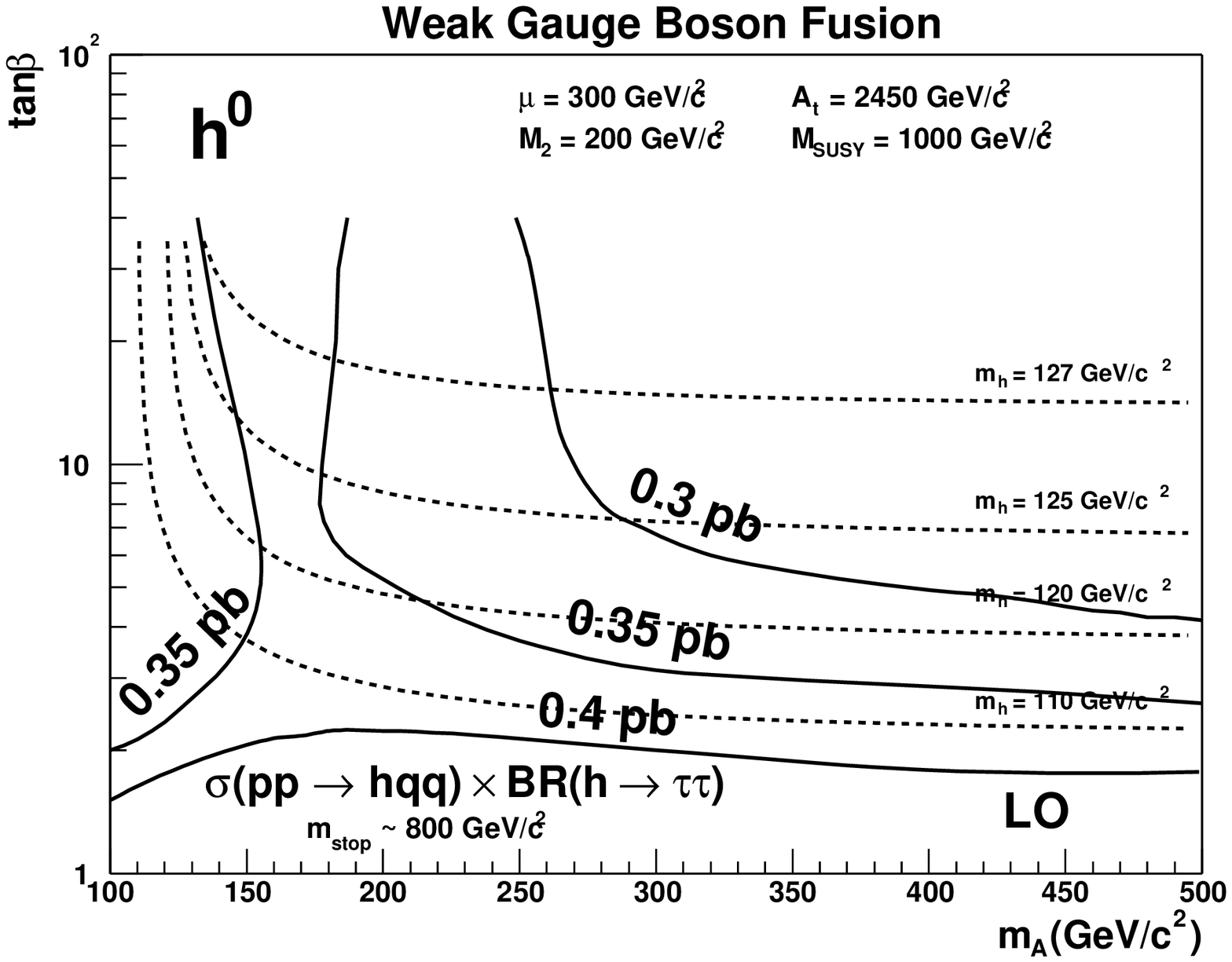}
  \end{minipage}
  &
  \begin{minipage}{7.5cm}
    \centering
    \includegraphics[height=75mm,width=80mm]{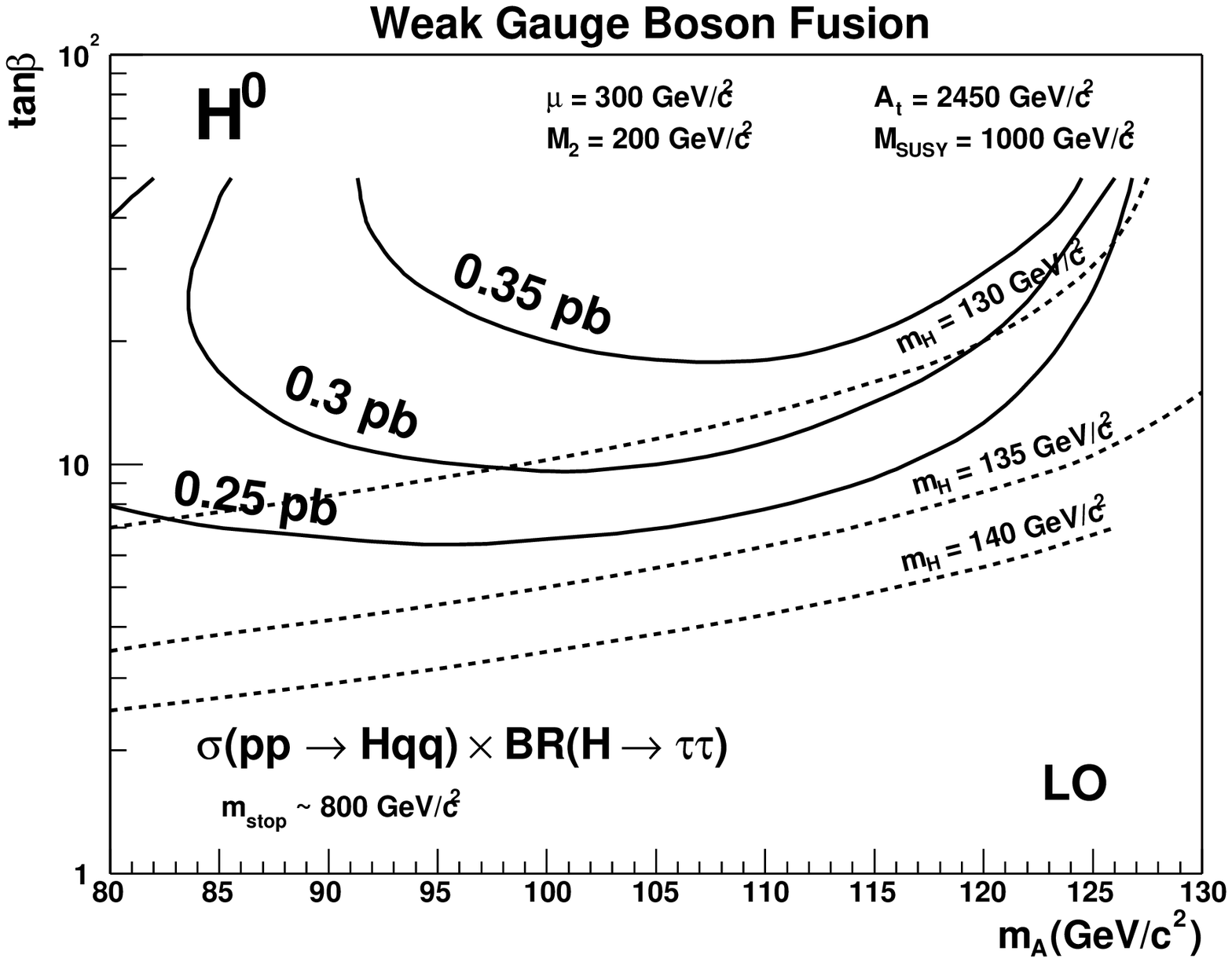}
  \end{minipage}\\
  \begin{minipage}{7.5cm}
    \centering
    \caption{Isorate curves for h$\rightarrow\tau^+\tau^-$ in the
             weak gauge boson fusion qq~$\ra$~qqh. Maximal stop
             mixing and LO cross sections are assumed.}
    \label{fig:hl_tautau}
  \end{minipage}
  &
  \begin{minipage}{7.5cm}
    \centering
    \caption{Isorate curves for H$\rightarrow\tau^+\tau^-$ in the
             weak gauge boson fusion qq~$\ra$~qqH in the region of
             the (m$_{\rm A},\tan\beta$) parameter space where H
             is SM-like. Maximal stop mixing and LO cross sections
             are assumed.}
    \label{fig:hh_tautau}
  \end{minipage}
  \end{tabular}
\end{figure}

The SM Higgs boson is expected to be accessible in the weak gauge boson 
fusion production process 
$\rm qq \ra \rm qqH$ for m$_{\rm H}\lesssim$~150~GeV/$c^2$ with the
$\rm H \ra \gamma\gamma$ \cite{dubinin}, 
$\rm H \ra \rm WW^* \ra \ell^+\ell^-\nu_{\ell}\nu_{\ell}$ \cite{qqh_ww} and 
$\rm H \ra \tau^+\tau^-$  \cite{sasha} decay channels
for integrated luminosities exceeding $\sim$~60~$\rm fb^{-1}$.
For the $\rm H \ra \gamma \gamma$ decay channel the total rate required 
with 60~$\rm fb^{-1}$ is about 8~fb for 
$\rm m_{\rm H}$~=~115 GeV/$c^2$ and 6.6~fb for $\rm m_{\rm H}$~=~127~GeV/$c^2$ 
\cite{dubinin}.
The $\rm H \ra \tau^+\tau^-$ channel has been studied with lepton-plus-jet final states \cite{sasha}.
 The total rate required
for a 5$\sigma$ statistical significance with an integrated luminosity of 
30~$\rm fb^{-1}$ varies from 0.4 to 0.28~fb for
115~$<\rm m_{\rm H}<$~127~GeV/$c^2$ and from 0.28 to~0.19 fb in the interval of 
127~$<\rm m_{\rm H}<$~145~GeV/$c^2$ for the searches of the SM-like heavy scalar H. 
Figures \ref{fig:wwhgg} and \ref{fig:hl_tautau} show the isorate curves for the $\rm h \ra \gamma \gamma$ 
and $\rm h \ra \tau^+\tau^-$ decay channels in the weak gauge boson fusion process
in the maximal-mixing scenario with LO cross sections. 
Figure \ref{fig:hh_tautau} shows the corresponding isorate curves for the heavy scalar
MSSM Higgs boson H in the $\rm H \ra \tau^+\tau^-$ decay channel.
As can be seen
seen from Fig.~\ref{fig:hl_tautau}, the $\rm h \ra \tau^+\tau^-$ channel could be accessible in 
a large part of the parameter space already with low integrated luminosities. A sensitivity
at large $\rm m_{\rm A}$ and tan$\beta$ is expected also in the 
$\rm h \ra \rm WW^* \ra \ell^+\ell^-\nu_{\ell}\nu_{\ell}$ 
decay channel because the studies in the SM framework indicate a 5$\sigma$ discovery for 
$\rm m_{\rm H} \gsim$~120~GeV/$c^2$ \cite{qqh_ww}. 

\section{Discovery potential}\label{sec:discovery_ranges}

\pagestyle{empty}
\begin{figure}[p]
  \centering
  \includegraphics[height=100mm,width=150mm]{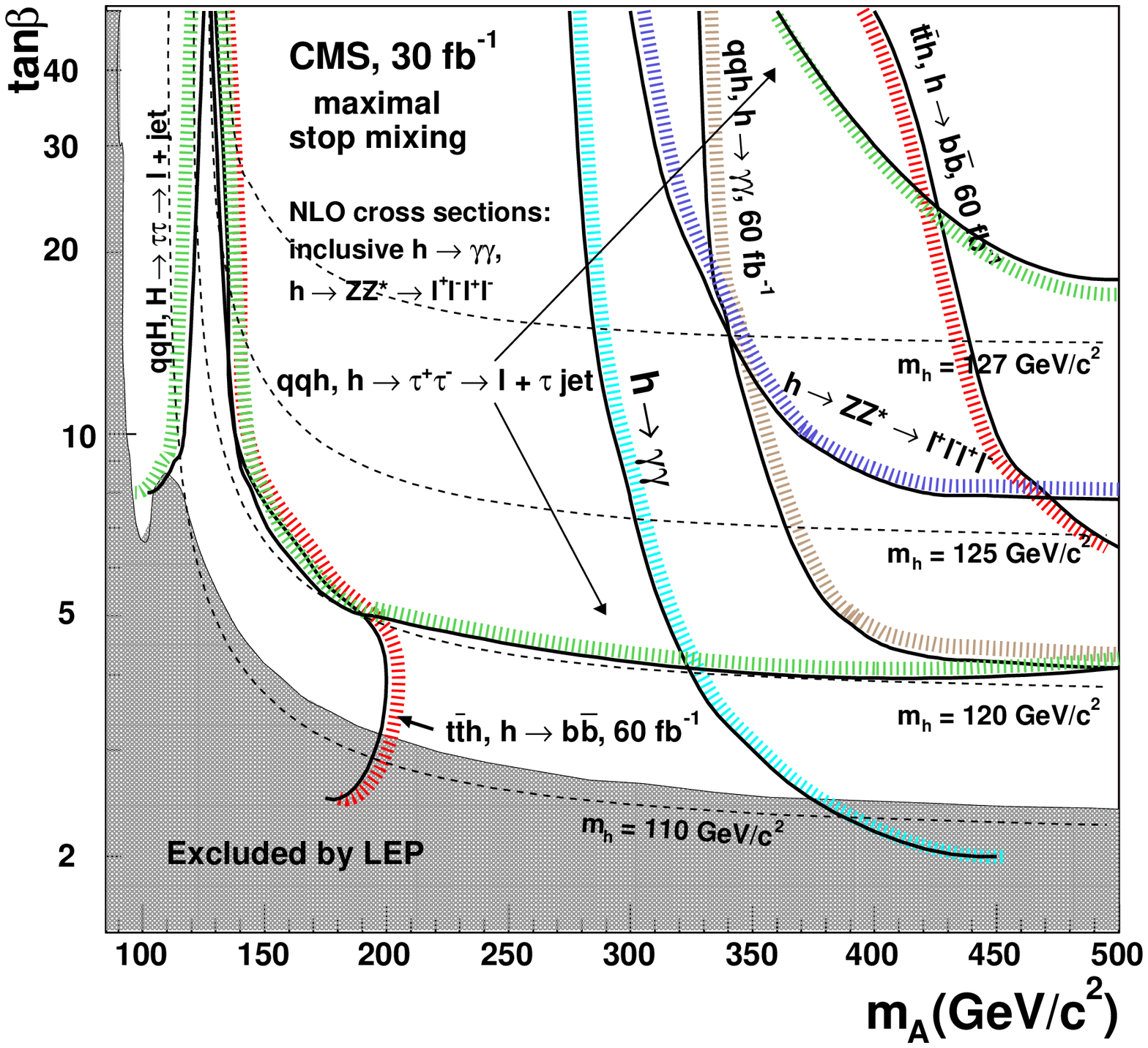}
  \caption{The 5$\sigma$-discovery potential of CMS for the lightest MSSM Higgs
  boson as a function of $\rm m_{\rm A}$ and tan$\beta$ for 30~fb$^{-1}$ with
  maximal stop mixing. The reach in the
  $\rm qq \ra \rm qqh$, $\rm h \ra \gamma\gamma$ channel 
  is shown for 60~fb$^{-1}$. The discovery potential for the $\rm t\overline{\rm t}\rm h$,
  $\rm h \ra$ b$\overline{\rm b}$ channel
  for 60~fb$^{-1}$ is taken from Ref.~\cite{volker2}. The reach of the 
  $\rm H \ra \tau^+\tau^-$ decay channel of the heavy scalar in the $\rm qq \ra \rm qqH$
  production process is also shown.}
  \label{fig:5sigma30fb}
  \centering
  \includegraphics[height=100mm,width=150mm]{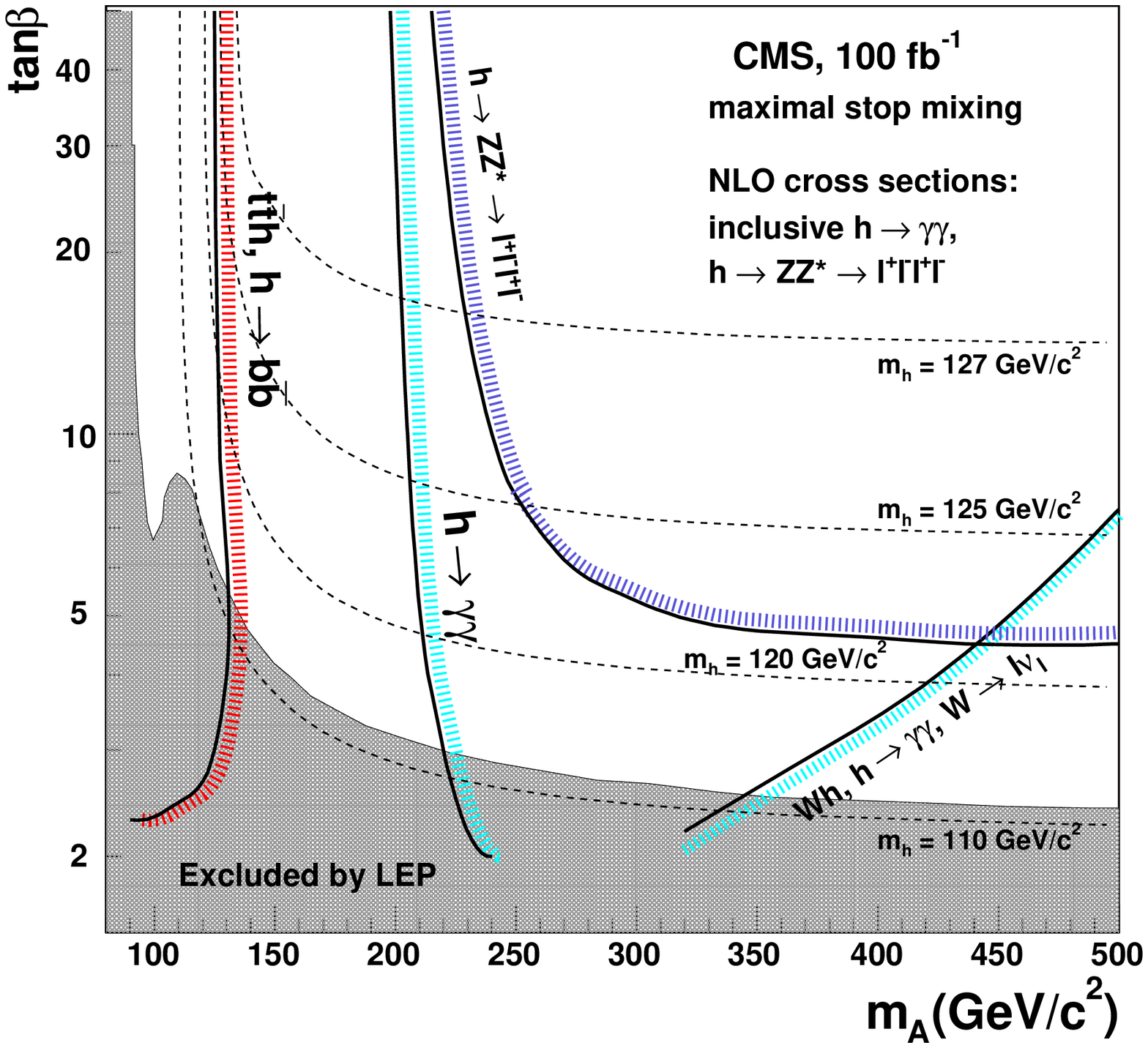}
  \caption{The 5$\sigma$-discovery potential for the lightest MSSM Higgs
  boson as a function of $\rm m_{\rm A}$ and tan$\beta$ for 100~fb$^{-1}$ with
  maximal stop mixing. The discovery potential for the $\rm t\overline{\rm t}\rm h$,
  $\rm h \ra$ b$\overline{\rm b}$ channel is taken from Ref.~\cite{volker2}.}
  \label{fig:5sigma100fb}
\end{figure}

Figures \ref{fig:5sigma30fb} and \ref{fig:5sigma100fb}
show the discovery potential of CMS for the lightest MSSM
Higgs boson as a function of $\rm m_{\rm A}$ and tan$\beta$ assuming maximal stop mixing,
$\rm m_{\rm top}$~=~175~GeV/$c^2$ and $\rm M_{\rm SUSY}$~=~1~TeV/$c^2$, for 30~$\rm fb^{-1}$ 
and  100~$\rm fb^{-1}$, respectively.
The $\rm h \ra \gamma\gamma$ and
 $\rm h \ra \rm ZZ^* \ra \ell^+\ell^-\ell^{\prime +}\ell^{\prime -}$ decay channels
in the inclusive production are shown with NLO cross sections. 
With an integrated luminosity of 100~$\rm fb^{-1}$ these
channels cover a major part of the MSSM parameter space, for $\rm m_{\rm A}\gsim$~200~GeV/$c^2$
and $\rm m_{\rm A}\gsim$~250~GeV/$c^2$, respectively. 
In the associated $\rm qq \ra \rm Wh$ production the $\rm h \ra \gamma\gamma$ channel
covers only a small region at
large $\rm m_{\rm A}$ and small ($\lesssim$~5) tan$\beta$ values. The
sensitivity for the $\rm h \ra$ b$\overline{\rm b}$ channel in the associated 
$\rm q\overline{\rm q}/ \rm gg \ra \rm t\overline{\rm t}\rm h$ production with 
60~$\rm fb^{-1}$ from Ref.~\cite{volker2}
is also shown in the figure.
The reach in the $\rm h \ra \gamma\gamma$ and $\rm h \ra \tau^+\tau^-$ decay channels in the
weak gauge boson fusion production is
shown in Fig.~\ref{fig:5sigma30fb} for 60 and 30~$\rm fb^{-1}$, respectively. 
In this production mode, the region tan$\beta \gsim$~5 can be covered
with the $\rm h \ra \gamma\gamma$ channel for
$\rm m_{\rm A}\gsim$~350~GeV/$c^2$ and with the $\rm h \ra \tau^+\tau^-$ channel 
for $\rm m_{\rm A}\gsim$~120~GeV/$c^2$.
The $\rm H \ra \tau^+\tau^-$ decay channel of the heavy scalar, shown also in
 Fig.~\ref{fig:5sigma30fb}, covers the region
 $\rm m_{\rm A}\lesssim$~125~GeV/$c^2$ in the weak gauge boson fusion. 
The region 90~GeV/$c^2 \lesssim \rm m_{\rm A}\lesssim$~130~GeV/$c^2$ at large tan$\beta$,
where the lightest Higgs boson is no more SM-like, is
outside the reach of the channels discussed in this paper.  
To explore this region, the $\rm h \ra \mu^+\mu^-$ and $\rm h \ra \tau^+\tau^-$ decay channels
can be used in the associated production with b quarks,
$\rm q\overline{\rm q}/\rm gg \ra \rm b \overline{\rm b}\rm h$, exploiting the enhanced
couplings to down type fermions in the MSSM at large tan$\beta$.

\vspace{ 3mm}
\section{Conclusions}
\vspace{ 3mm}

The production of the lightest MSSM Higgs boson h was studied, effects of 
SUSY parameters were discussed.
The discovery potential was evaluated for CMS in the maximal-mixing scenario
for the inclusive $\rm  h \ra\gamma\gamma$ channel, for the $\rm  h \ra\gamma\gamma$
channel in the associated production Wh and $\rm t\overline{\rm t}\rm h$, for 
the $\rm h \ra \tau^+\tau^-$ channel in the weak gauge boson fusion and,
for the first time, for the $\rm h \ra \rm ZZ^* \ra \ell^+\ell^-\ell^{\prime +}\ell^{\prime -}$
channel at large tan$\beta$. Consequencies of a light stop quark were shown
for the expected discovery regions.
 
Already with an integrated luminosity of 30~fb$^{-1}$ the parameter space 
$\rm m_{\rm A}\gsim$~150~GeV/$c^2$ and tan$\beta \gsim$~5,
apart from a small region at large $\rm m_{\rm A}$ and tan$\beta$,
 is covered with the 
$\rm h \ra \tau^+\tau^-$ decay channel in the weak gauge boson fusion qq~$\ra~\rm qqh$.
The reach with the $\rm  h \ra\gamma\gamma$ decay channel with 30~fb$^{-1}$ 
is for $\rm m_{\rm A}\gsim$~300~GeV/$c^2$
in the inclusive production and for $\rm m_{\rm A}\gsim$~350~GeV/$c^2$ in the 
$\rm qq \ra qqh$ production process.
With 60~fb$^{-1}$ the parameter space 150~$\lesssim \rm m_{\rm A}\lesssim$~400~GeV/$c^2$
 ($\rm m_{\rm A}\gsim$~150~GeV/$c^2$ for tan$\beta \lesssim$~5)
 is covered with the 
$\rm h \ra\rm b \overline{\rm b}$ decay channel in the
$\rm q\overline{\rm q}/\rm gg \ra \rm t\overline{\rm t}\rm h$ production process. 
With the large integrated luminosity of 100~fb$^{-1}$ 
the inclusive $\rm  h \ra\gamma\gamma$ channel yields a 
5$\sigma$-discovery for $\rm m_{\rm A}\gsim$~200~GeV/$c^2$ and the 
$\rm h \ra \rm ZZ^* \ra \ell^+\ell^-\ell^{\prime +}\ell^{\prime -}$ channel
 for $\rm m_{\rm A}\gsim$~250~GeV/$c^2$, tan$\beta \gsim$~5. 

The effects of loop corrections to the cross sections and branching ratios
were studied in a scenario with large mixing and light stop quark. The 
consequences of the stop-top interference effects were shown for the  
$\rm h \ra\gamma\gamma$ decay channel in the gluon fusion and in the inclusive 
production. The reduction 
of the total production rate was found to be significant for 
$\rm m_{\rm stop}\lesssim$~300~GeV/$c^2$. For 
$\rm m_{\rm stop}\lesssim$~200~GeV/$c^2$ the 
sensitivity in the inclusive $\rm  h \ra\gamma\gamma$ channel could be entirely lost.
In this scenario the production rate
is slightly enhanced for the associated production processes 
$\rm q\overline{\rm q}/\rm gg \ra \rm t\overline{\rm t}\rm h$ and $\rm qq \ra \rm W \rm h$ 
and in the weak gauge boson fusion $\rm qq \ra qqh$ process due to the 
positive interference effects on the $\rm h \ra \gamma\gamma$ decay width.

\section{Acknowledgments}

The authors would like to thank Michael Spira for helpful comments and for his efforts 
in developing the program HIGLU compatible with the other programs used in this work.
P.S. and S.L. would also like to thank Katri Huitu for helpful discussions.

\pagestyle{plain}

\end{document}